\documentclass[longauth]{aa}  
\usepackage{xcolor}
\usepackage{graphicx}
\usepackage{txfonts}
\usepackage{subcaption}         
                                
\usepackage{lscape}             
                                
\usepackage{placeins}

\usepackage{natbib}
\bibpunct{(}{)}{;}{a}{}{,}

\usepackage[colorlinks = true,
            linkcolor = blue,
            urlcolor  = blue,
            citecolor = blue,
            anchorcolor = blue]{hyperref}

\usepackage{multirow}
\newcommand{\espresso}{ESPRESSO}
\newcommand{\HARPS}{HARPS}
\newcommand{\serval}{\texttt{SERVAL}}
\newcommand{\sbart}{\texttt{s-BART}}
\newcommand{\arve}{\texttt{ARVE}}

\newcommand{\tauCeti}{$\tau$ Ceti}
\newcommand{\epsIndi}{$\epsilon$ Indi}
\newcommand{\HARPSasteroStar}{HD160691}

\newcommand{\kilometersecond}{${\rm km}\ {\rm s}^{-1}$}
\newcommand{\metersecond}{${\rm m}\ {\rm s}^{-1}$}
\newcommand{\metersecondhour}{\hbox{${\rm m}\ {\rm s}^{-1}\ {\rm h}^{-1}$}}
\newcommand{\centimetersecond}{${\rm cm}\ {\rm s}^{-1}$}

\usepackage{orcidlink}
\usepackage{float}
\begin{document}
\nolinenumbers
\title{A systematic bias in template-based RV extraction algorithms}

\author{A. M. Silva\inst{\ref{inst1}, \ref{inst2}} \orcidlink{0000-0003-4920-738X}
	\and N. C. Santos \inst{\ref{inst1}, \ref{inst2}}
	\and J. P. Faria \inst{\ref{inst4}}
	\and J. H. C. Martins \inst{\ref{inst1}}
	\and E. A. S. Cristo \inst{\ref{inst1}}                   
	\and S. G. Sousa \inst{\ref{inst1}}
	\and P. T. P. Viana \inst{\ref{inst1}, \ref{inst2}}
	\and É. Artigau     \inst{\ref{inst6}, \ref{inst7}}			\orcidlink{0000-0003-3506-5667}
	\and K. Al Moulla \inst{\ref{inst1}}\thanks{SNSF Postdoctoral Fellow} \orcidlink{0000-0002-3212-5778}
	\and A. Castro-González \inst{\ref{inst16}}			\orcidlink{0000-0001-7439-3618}		
	\and D. F. M. Folha   \inst{\ref{inst1}, \ref{inst5}}  		 \orcidlink{0000-0003-0150-0288}
	\and P. Figueira \inst{\ref{inst1}, \ref{inst4}}				 \orcidlink{0000-0001-8504-283X}
	\and T. Schmidt\inst{\ref{inst4}}
	\and F. Pepe\inst{\ref{inst4}}     									 
	\and X. Dumusque\inst{\ref{inst4}}     						 
	\and O. D. S. Demangeon   \inst{\ref{inst1}}				 
	\and T. L. Campante\inst{\ref{inst1}, \ref{inst2}}     		\orcidlink{0000-0002-4588-5389}
	\and X. Delfosse	\inst{\ref{inst11}}								
	\and B. Wehbe\inst{\ref{inst3}, \ref{inst15}}  				 
	\and J. Lillo-Box		\inst{\ref{inst16}}		
	\and A. R. Costa Silva \inst{\ref{inst1}, \ref{inst2}, \ref{inst4}}		\orcidlink{0000-0003-2245-9579}
	\and J. Rodrigues\inst{\ref{inst1}, \ref{inst2}, \ref{inst18}}			\orcidlink{0000-0001-5164-3602}
	\and J. I. Gonz\'alez Hern\'andez 	\inst{\ref{inst8}, \ref{inst9}}		\orcidlink{0000-0002-0264-7356}
	\and T. Azevedo Silva 	\inst{\ref{inst19}}								\orcidlink{0000-0002-9379-4895}				
	\and S. Cristiani		\inst{\ref{inst10}}	
	\and H. M. Tabernero \inst{\ref{inst12}, \ref{inst13}}		
	\and E. Palle 	\inst{\ref{inst9}, \ref{inst8}}  			\orcidlink{0000-0003-0987-1593}
	\and B. Lavie \inst{\ref{inst4}}   \orcidlink{0000-0001-8884-9276}
	\and A. Su\'arez Mascare\~no \inst{\ref{inst8}, \ref{inst9}}	\orcidlink{0000-0002-3814-5323}
	\and P. Di Marcantonio \inst{\ref{inst10}}   \orcidlink{0000-0003-3168-2289}
	\and C. J. A. P. Martins \inst{\ref{inst1}, \ref{inst17}}
	\and N. J. Nunes \inst{\ref{inst3}}   \orcidlink{0000-0002-3837-6914}
	\and A. Sozzetti \inst{\ref{inst20}} \orcidlink{0000-0002-7504-365X}
}
  
\institute{
	Instituto de Astrof\'{\i}sica e Ci\^encias do Espa\c co, CAUP, Universidade do Porto, Rua das Estrelas, 4150-762 Porto, Portugal\label{inst1}
	\and 
	Departamento de F\'{\i}sica e Astronomia, Faculdade de Ci\^encias, Universidade do Porto, Rua do Campo Alegre, 4169-007 Porto, Portugal\label{inst2} 
	\and
	Instituto de Astrof\'{\i}sica e Ci\^encias do Espa\c{c}o, Universidade de Lisboa, Campo Grande, 1749-016 Lisboa, Portugal\label{inst3}
	\and 
	Département d’astronomie de l’Universit\'e de Gen\`eve, Chemin Pegasi 51, 1290 Versoix, Switzerland \label{inst4}
	\and
	University Institute of Health Sciences, CESPU, CRL, 4585-116 Gandra, Portugal \label{inst5}
	\and 
	Institut Trottier de recherche sur les exoplanètes, Université de Montréal, 1375 Ave Th\'rèse-Lavoie-Roux, Montr\'al, QC, H2V 0B3, Canada \label{inst6}
	\and  
	Observatoire du Mont-M\'egantic, Universit\'e de Montr\'eal, Montr\'eal H3C 3J7, Canada\label{inst7}
	\and 
	Instituto de Astrof{\'\i}sica de Canarias, E-38205 La Laguna, Tenerife, Spain \label{inst8}
	\and
	Universidad de La Laguna, Dept. Astrof{\'\i}sica, E-38206 La Laguna, Tenerife, Spain \label{inst9}
	\and INAF-Astronomical Observatory.  via Tiepolo 11, 34143 Trieste, Italy \label{inst10}
	\and Univ. Grenoble Alpes, CNRS, IPAG, 38000 Grenoble, France\label{inst11}
	\and Institut de Ciències de l’Espai (ICE, CSIC), Campus UAB, c/ de Can Magrans s/n, 08193 Cerdanyola del Vallès, Barcelona, Spain \label{inst12}
	\and Institut d'Estudis Espacials de Catalunya (IEEC), Edifici RDIT, Campus UPC, 08860 Castelldefels (Barcelona), Spain \label{inst13}
	\and Departamento de Física, Faculdade de Ciências, Universidade de Lisboa, Campo Grande, P-1749-016 Lisboa, Portugal \label{inst15}
	\and Centro de Astrobiolog{\'\i}a (CAB), CSIC-INTA, Camino Bajo del Castillo s/n, 28692, Villanueva de la Ca\~nada, Madrid, Spain \label{inst16}
	\and Centro de Astrof\'{\i}sica da Universidade do Porto, Rua das Estrelas,	4150-762 Porto, Portugal \label{inst17}
	\and Observatoire François-Xavier Bagnoud -- OFXB, 3961 Saint-Luc, Switzerland \label{inst18}
	\and INAF – Osservatorio Astrofisico di Arcetri, Largo Enrico Fermi 5, 50125 Firenze, Italy, \label{inst19}
	\and INAF – Osservatorio Astrofisico di Torino, Via Osservatorio 20, 10025 Pino Torinese, Italy, \label{inst20}
	}
   
\date{Received September 30, 20XX}

\abstract
{The radial velocity (RV) method plays a key role in modern-day astrophysics. One of the most common techniques to extract precise RVs from state-of-the-art spectrographs is template-matching (TM) algorithms. They have been shown to perform better than cross correlation (CCF) methods in cooler stars (e.g., M dwarfs) and multiple implementations have appeared over the past years. More recently, line-by-line (LBL) approaches offer an alternative avenue to extract RVs by analyzing individual spectral lines.}
{In this paper we identify and explore a previously unidentified, multi meter-per-second, systematic correlation between time and RVs inferred through TM and LBL methods. We evaluate the influence of the data-driven stellar template in the RV bias and hypothesize on the possible sources of this effect.}
{We first use the \sbart{} pipeline to extract RVs from three different datasets gathered over four nights of \espresso{} and \HARPS{} observations. Then, we demonstrate that the effect can be recovered on a larger sample of 19 targets, totaling 4124 \espresso{} observations spread throughout 38 nights. We also showcase the presence of the bias in RVs extracted with the \serval{} and \arve{} pipelines. Lastly, we explore the construction of the stellar template through the 5 years of \espresso{} observations of HD10700, totalling more than 2000 observations.}
{We find that a systematic quasi-linear bias affects the RV extraction with slopes that vary from $\sim$ -0.3 \metersecondhour to $\sim$ -52 \metersecondhour{} in our sample. This trend is not observed in CCF RVs and appears when all observations of a given star are collected within a short time-period (timescales of hours). We show that this systematic contamination exists in the RV time-series of two different template-matching pipelines, one line-by-line pipeline, and that it is agnostic to the spectrograph. We also find that this effect is connected with the construction of the stellar template, as we are able to mitigate it through a careful selection of the observations used to construct it. Our results suggest that a contamination of micro-telluric features, coupled other sources of correlated noise, could be the driving factor of this effect. We also show that this effect does not impact the usual usage of template-matching for the detection and characterization of exoplanets. Other short-timescale science cases -- such as, asteroseismology, transit and atmospheric characterization -- can however be severely affected.}
{}


\maketitle

\section{Introduction} 

The field of exoplanets has seen a massive growth over the past three decades, with 5849\footnote{Information from March 11th, 2025; Refer to \url{https://exoplanetarchive.ipac.caltech.edu} for updated number.} detections of exoplanetary signals. Despite this impressive growth, the discovery and characterization of an "Earth-analog" still remains a major challenge, driving improvements in both instrumentation and data analysis techniques. A major work-horse of the field is the radial velocity (RV) method, based on the measurement of variations of a star's velocity along the line of sight. This is done through the measurement of displacements in the location of the spectral lines, in comparison to their expected position. 

Often, RV extraction is done through the Cross Correlation Function (CCF) method -- based on the cross-correlation between stellar spectra and a (weighted) binary mask \citep{baranne_elodie_1996,pepe_coralie_2002} -- but this is not always the optimal approach. In cooler stars (e.g., M dwarfs) the template matching (TM) algorithm typically yields better results, leading to smaller RV scatters and improved RV precision, as seen in different implementations, e.g, \texttt{HARPS-TERRA} \citep{anglada_escude_HARPS_TERRA_2012}, \texttt{NAIRA} \citep{astudillo-defru_search_2015}, and \texttt{SERVAL} \citep{zechmeisterSpectrumRadialVelocity2018}.	Template matching is a data-driven method, based on the construction of a high-SNR stellar model from available observations of a given star, which is then aligned with the individual observations. Usually, TM methods share a common set of characteristics: i) chromatically-independent RV shifts, where each spectral order is assumed to be an independent measurement of the same underlying phenomenon; ii) the stellar template (model) is capable of accurately describing every observation of a given star. \sbart{}\footnote{Publicly availabe at \url{https://github.com/iastro-pt/sBART}} \citep{silvaNovelFrameworkSemiBayesian2022} tackled the assumption of chromaticity, as the RV signal of exoplanets is independent of the wavelength at which it is measured. This is done by imposing a wavelength-independent RV shift -- common to all spectral orders -- to describe the differences between the stellar model and observations. Since its inception, \sbart{} has used for the discovery and characterization of several systems across  different state-of-the-art spectrographs \citep[e.g.,][]{fariaCandidateShortperiodSubEarth2022, palethorpe_confronting_2024, balsalobre-ruza_kobe-1_2025}. In order to refine template matching avenues we have seen a push towards a careful selection of the lines that are used for RV extraction. The line-by-line (LBL) methods \citep[e.g.,][]{dumusque_measuring_2018,artigau_line-by-line_2022} share a common working principle with TM methods: a high-SNR stellar model is constructed from the data and, afterwards, RVs are estimated through their analysis. This process is carried out for each individual spectral line and, at the end, a selection is made in order to remove the ones that present a larger variability.

In exoplanetary RV blind-searches, or in the follow-up of transiting exoplanets, observations are typically carried out over multiple months. This allows to properly sample the orbital period of the orbiting companions and to characterize any stellar signals. However, some science cases require a much shorter cadence of observations, such as transmission or emission spectroscopy studies \citep[e.g., ][]{wyttenbach_spectrally_2015, costa_silva_espresso_2024}, or asteroseismology analysis \citep[e.g,][]{bouchy_asteroseismology_2005,campante_expanding_2024}. Similarly, the RV detection of planets with unusually ultra-short periods \citep[e.g, $P_{\rm orb}$ $<$ 12 h,][]{adams_ultrashortperiod_2021, castrogonzalez_two_2025} can be achieved through very short observing baselines. These datasets are typically collected within a single night, or over a few consecutive nights. 

\textit{A priori}, one would expect the RV extraction through a template-based algorithm -- TM or LBL -- to perform similarly for different observing strategies. However, in this paper we show that this is not necessarily the case. In particular, we present a systematic bias in the RV extraction through template-based algorithms when the stellar model is constructed from observations collected within a short time-span, e.g., within a night. This bias appears in two TM pipelines, one LBL pipeline, is present in the data of multiple state-of-the-art spectrographs, and it has different amplitudes for different stars. The bulk of our analysis is done mainly through the publicly available \sbart{} pipeline, using data from \espresso{} \citep{pepeESPRESSOVLTOnsky2021} and \HARPS{} \citep{2003Msngr.114...20M,pepe_harps_2002}.

We start by providing a description of our datasets in Section \ref{App:dset_overview}, followed by an overview of the methods for RV extraction in Section \ref{Sec:dset_and_sbart}. Then, in Section \ref{Sec:trend_introduction} we showcase the magnitude of this spurious signal in different stars and explore its wavelength-dependence. In Section \ref{Sect:tauceti_night_by_night} we focus our analysis on \espresso{} observations and explore the effect that the selection of observations has on the retrieved RVs. In Section \ref{Sec:the_hypothesis_of_tellurics} we hypothesize on the root cause of this systematic bias. Lastly, we wrap-up in Section \ref{Sec:discussion} with a discussion on the possible avenues to mitigate this bias, as well as the impact it introduces on different science cases.

\section{Overview of the observations} \label{App:dset_overview}

	We use observations collected with the \espresso{} and \HARPS{} spectrographs. The raw data was processed with the corresponding version of the official data reduction software (DRS) of each spectrograph\footnote{Available at \url{https://www.eso.org/sci/software/pipe_aem_table.html}}. It handles the extraction of spectral orders from the raw images, performs the necessary calibrations and corrections, and provides an RV extraction through the CCF method. The CCF RVs are obtained through a Gaussian fit to the outcome of cross-correlating the stellar spectra with a weighted mask that is selected from a pre-defined list. This selection is based on the spectral type, using the one that most resembles the star that is being processed. We used \HARPS{} DRS version \textit{3.2.5}, whilst the bulk of \espresso{} data was reduced with pipeline version \textit{3.0.0}\footnote{We note that the results from this paper do not present measurable differences when using spectra reduced with different pipeline versions.}.

	Our analysis will be focused on three different datasets: asteroseismology observations, long-period coverage of \tauCeti, and a large sample of single-night observations. We provide a detailed description of each dataset in the next sub-sections.

	\subsection{Asteroseismology datasets} \label{App:dset-overview_astero}

		Our asteroseismology datasets are comprised of observations of three different stars:

		\begin{enumerate}
			\item \textit{HD160691} - This G3 star has 2183 \HARPS{} public spectra, available through the \textit{DACE} interface, under \textit{ESO} program \textit{073.D-0578(A)}. This dataset was the basis for the discovery of the first Neptune-like exoplanet ever announced \citep{santos_harps_2014}. For our analysis we are only interested in the asteroseismology time-series (i.e., those that span a few hours). We found 6 consecutive nights \citep[2004-06-05 to 2004-06-10,][]{bouchy_asteroseismology_2005} from which we select the last two for our analysis. 
			\item \textit{HD 209100 (\epsIndi{})} - This K5 star was observed with \espresso{} for 6 nights on December 2022 -- \textit{ESO} program \textit{109.236P.001} -- allowing the detection of solar-like oscillations. These have a peak amplitude (radial modes) of 2.6 \centimetersecond, making this the coolest seismic dwarf ever observed  \citep{campante_expanding_2024}. 
			\item \textit{HD40307} - This bright K-dwarf star was observed during a 5-night campaign during \espresso{}'s commissioning  -- \textit{ESO} program \textit{0102.D-0346(A)} -- with a tentative detection of p-mode oscillations \citep{pepeESPRESSOVLTOnsky2021}. However, the dataset was revisited by \citet{campante_expanding_2024} which found no evidence for their presence.
		\end{enumerate}
		
		In total we used 475 \HARPS{} observations of \HARPSasteroStar{}, 2084 \espresso{} observations of \epsIndi{}, and 1140 \espresso{} observations of HD40307. The BERV coverage of the three stars can be found in Figure \ref{fig:astero_berv_over_night_app}, and Table \ref{Tab:astero_snr_over_night_app} presents an overview of the nights in which the stars were observed, alongside their median SNR at a wavelength of 600 nm. 

	\subsection{HD10700 (\tauCeti)} \label{Sec:dset_overview_tauCeti}

		HD10700, commonly referred to as \tauCeti{}, is a bright G8 star that was intensively observed within the Guaranteed Time of Observations (GTO) of \espresso\footnote{Guaranteed Time Observations collected at the European Southern Observatory under \textit{ESO} program IDs:  110.24CD.001, 1102.C-0744(X), 1104.C-0350(X), 1104.C-0350(Q), 1102.C-0744(O), 108.2254.006, 108.2254.001, 1102.C-0744(N), 60.A-9129(A), 1104.C-0350(R), 1104.C-0350(B), 110.24CD.003, 110.24CD.009, 60.A-9128(G), 108.2254.004, 1102.C-0744(B), 1102.C-0744(M), 1104.C-0350(P), 108.2254.003, 1104.C-0350(D), 106.21M2.001, 1104.C-0350(Y)}. The \espresso{} spectrograph observed \tauCeti{} for close to 5 years with a median number of 15 observations per night. In total, 2442 observations were collected and a detailed analysis of this system (using CCF and \sbart{} RVs) can be found in \citet{figueira_figueiraespressogto_nodate}. Unlike other \espresso{} observations, the \tauCeti{} dataset was reduced with \espresso{} DRS version \textit{3.3.0} for the analysis of \citet{figueira_figueiraespressogto_nodate}. Figure \ref{fig:TauCetiBERV} presents the BERV coverage of the \espresso{} observations, highlighting in blue the nights for which we find a decrease in BERV (in comparison with the previous night). In this paper we also looked at individual nights of this star, whose BERV distribution is shown in Figure \ref{Fig:tauceti_single_nights_bervdist} and some of their properties are summarized in Table \ref{Tab:tauceti_nights_summary}. 

	\subsection{Large sample of single-night observations} \label{Sec:dset_overview_large_sample}
		
		The last dataset in our analysis is comprised of a large sample of \espresso{} targets. The majority of them were selected from within \espresso{} \textit{GTO}'s working group 2 - \textit{Atmosphere characterization}, as it collected a large number of multi-hour time-series over multiple nights\footnote{Guaranteed Time Observations collected at the European Southern Observatory under \textit{ESO} program IDs: 073.D-0578(A), 0102.D-0346(A), 108.2254.006, 109.236P.001, 1104.C-0350(K), 1104.C-0350(A), 1104.C-0350(L), 1102.C-0744(O), 110.24CD.003, 1104.C-0350(I), 1104.C-0350(D), 1102.C-0744(V), 1104.C-0350(J), 1104.C-0350(Q), 106.21M2.004, 1104.C-0350(V), 1104.C-0350(F), 108.2254.003}. All in-transit observations were discarded from the analysis to avoid contamination from the spectral imprint of a transiting exoplanet. This is done through a manual identification of the observations taken outside of the planetary transit (typically known as "out-of-transit" observations). To those observations, we add the \espresso{} observations of \textit{HD40307} and \epsIndi{} (previously described in Section \ref{App:dset-overview_astero}).

		In total, we collected 4124 \espresso{} observations of 19 different stars that were observed over 38 nights. Figure \ref{fig:largesample_EBRVdist} presents an histogram of the BERV distribution within this sample, showing a good coverage of the $\pm$ 30 \kilometersecond{} maximum Barycentric Earth Radial Velocity (BERV) variation over the year.

\section{RV extraction methods} \label{Sec:dset_and_sbart}

	In this section we present an overview of the RV-extraction process of the different algorithms that will be applied to the stellar spectra. In particular, we focus on the \sbart{} algorithm, as it will be used for the bulk of our analysis.

	\subsection{The s-BART algorithm for template-matching RVs} \label{APP:sbart_primer}
		
		The \textit{Semi-Bayesian Approach for RVs with Template matching} (\sbart{}) algorithm \citep{silvaNovelFrameworkSemiBayesian2022} is a TM algorithm for RV extraction, built around a common assumption: the RV shift that an orbiting companion introduces is independent of the wavelength at which it is measured. This is enforced by assuming a common RV shift to describe all differences between the stellar model (the template) and the data with which the template is compared to (the individual spectra). Furthermore, by implementing this algorithm within a semi-Bayesian context, we have a straightforward and consistent method to characterize the posterior distribution of the RVs.

		We present here a brief overview of the RV extraction process, but we refer to \citet{silvaNovelFrameworkSemiBayesian2022} for a full description:

		\begin{itemize}
			\item We reject wavelength regions based on the quality control (QC) flags provided by the instrument's official pipeline. We also reject wavelength regions that we know to be sensitive to activity and reject spectral orders that do not meet internal quality control flags of \sbart.
			\item We construct a transmittance profile of Earth's atmosphere using \textit{Telfit} \citep{gulliksonCorrectingTelluricAbsorption2014}, assuming the worst atmospheric conditions of our dataset, defined as the observation with the highest relative humidity. Then, we construct a binary mask to reject any telluric line that has a depth greater than 1\%. This is a very conservative threshold, leading to a rejection of several wavelength regions. However, it also minimizes the number of micro-telluric features that are left in the stellar spectra. This binary mask is also expanded to account for the maximum yearly BERV variation of the star.
			\item Using all observations, we construct an high signal-to-noise-ratio (SNR) model by stacking the individual spectra and propagating the flux uncertainty. In this process, we place all observations in a common-rest frame using a previous RV estimate, e.g., computed through the CCF approach or an initial run of the TM algorithm. Then, we interpolate the spectra to a common wavelength grid. We also enforce that any wavelength gap in any of the observations will be translated into a gap in the stellar template, ensuring that every pixel in the stellar template has a corresponding pixel in each observation.
			\item We apply the \sbart{} algorithm to align the stellar template with the spectra. The RV posterior distribution caracterized through a \textit{Laplace} approximation, which assumes the posterior distribution to have a Gaussian shape. The RV measurement is obtained through the maximum of the posterior distribution, and the associated uncertainty through its standard deviation. During the RV extraction process we account for changes in the stellar continuum through the marginalization of a first-degree polynomial to describe differences between stellar spectra and template.
		\end{itemize}

		It is also important to note that in our RV extraction we ensure that we use the same orders in all observations. We note that this also translates into a loss of RV information, as the rejection of one order in one observation will imply the loss of that given order in all other observations. This, however, ensures that the final RV measurement of all observations will be informed by the same wavelength regions. 
	
	\subsection{The SERVAL pipeline for template-matching RVs}

		The \textit{SpEctrum Radial Velocity AnaLyser} (SERVAL) template matching pipeline is a widely used implementation of the template matching method. It constructs the stellar template through a B-spline regression of the continuum-normalized individual observations. Then, order-wise RVs are computed through a $\chi^2$ minimization between the Doppler-shifted template and the observations. Lastly, the order-by-order radial velocities are combined with a weighted mean to produce the final RV estimate for the observation.

	\subsection{The ARVE algorithm for line-by-line radial velocities}
		Analyzing Radial Velocity Elements\footnote{\url{https://github.com/almoulla/arve}} \citep[\arve ,][]{al_moulla_arve_nodate} is a Python code which can, among other functionalities, extract line-by-line (LBL) RVs. Unlike other LBL algorithms, and to speed up computations, \arve{} utilizes a grid of synthetic spectra, from which the properties of expected spectral lines within a provided wavelength range are retrieved. The LBL RVs are thereafter calculated through a per-line $\chi^2$-minimization between the observed spectrum and a template constructed from all or a subset of the observations. Specifically, as in \citet{bouchy_fundamental_2001}, the observed spectrum, $S_{\mathrm{obs}}$, is described as the scaled first-order Taylor expansion of the template spectrum, $S$,
		\begin{equation}
			S_{\mathrm{obs}} = A \left( S + \frac{\mathrm{d}S}{\mathrm{d}\lambda}\Delta\lambda \right) \,,
		\end{equation}
		in which the only free variables are the scaling, $A$, to account for integrated-flux differences between the observation and template, and the RV shift, $v$, which relates to the wavelength differential, $\Delta\lambda{=}{\lambda}v/c$.

\section{Systematic RV bias in short time baselines} \label{Sec:trend_introduction} 

\begin{figure}[H]
		\centering
		\resizebox{\hsize}{!}{\includegraphics{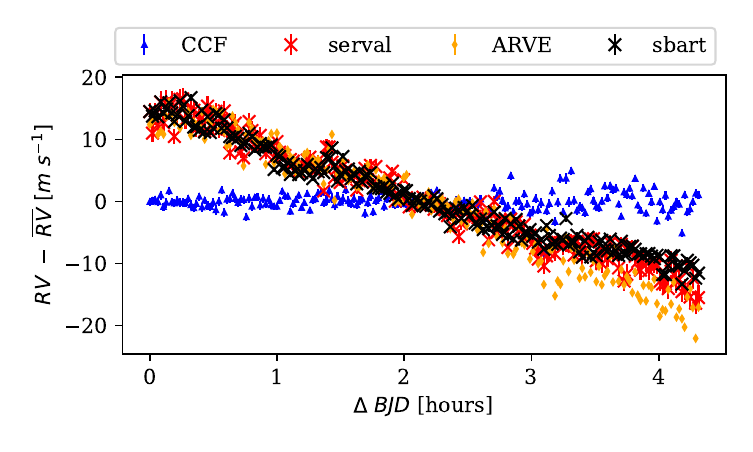}}
		\caption{Mean-subtracted RV time-series for one night (December 24, 2018) of \espresso{} observations of HD40307. The CCF RVs (blue triangles) are compared against those from \sbart{} (black), \serval{} (red), and \arve{} (orange).}
		\label{Fig:TREND:FIG:showcase}
	\end{figure}

	During the development and subsequent application of \sbart{} to multiple targets, we found a transversal issue to template-based techniques for RV extraction: if all observations are acquired within a short period (i.e., within a few days), then a quasi-linear systematic trend in RV appears over each individual night. This issue is present in the observations of multiple spectrographs and multiple stars, with varying amplitudes (Section \ref{Sect:large_sample_across_targets}). This RV bias was also reproduced using a different TM pipeline: the publicly available\footnote{\url{https://github.com/mzechmeister/serval}} version of \serval{}. In Figure \ref{Fig:TREND:FIG:showcase} we present a comparison of the RVs extracted for one night of \textit{HD40307}, showing that all template-based algorithms present a large peak-to-peak RV variation ($\sim$ 40 \metersecond), whilst the CCF RVs do not. 

	This rules out a possible pipeline error and reveals, instead, a systematic bias in template-matching RVs. This is, however, unexpected as \sbart{} -- similarly to other TM pipelines -- provides improvements on the cooler stars when the stellar template is constructed with data collected during a long time-series \citep[e.g.,][]{silvaNovelFrameworkSemiBayesian2022}. The fact that a LBL extraction of RVs also reveals a similar bias, leads to two possible explanations: 1) The issue is connected with the spectral lines, as LBL approaches do not use the continuum; or 2) The root source of the bias is linked to the stellar template, as both TM and LBL methods utilize a data-driven, high-SNR, stellar template. As shown in Appendix \ref{Sec:correlation_search_large_sample}, we fail to find a clear correlation between the RV residuals and the line-shape metrics that are derived from the CCF. In the next sections we will explore the behaviors of the observed trend.

	\subsection{Variation across targets} \label{Sect:large_sample_across_targets}
		
		In this section, we set out to characterize the behavior of this RV trend through a larger sample of observations (described in Section \ref{Sec:dset_overview_large_sample}). We will not include the \HARPS{} observations, preferring to use data from the same instrument, in order to avoid differences that could be caused by the instrument.
		
		We sub-divide the data of each star, so that each of the 38 nights of observations is treated fully independent from one another. Then, for each night, we apply the following prescription:
		
		\begin{enumerate}
			\item For each night we construct a telluric binary mask spanning the maximum yearly BERV variation for the target. This leads to a larger rejection of wavelength regions, but we prefer to keep a consistent telluric masking process across this paper.
			\item We apply the telluric mask to remove the telluric features of each observation of the night, which are then used to construct the stellar model and extract \sbart{} RVs.
			\item We compute the residuals between the \sbart{} and CCF RVs and fit a first degree polynomial to them. The fitting of a linear model is an simplification of the effect that we are measuring but, as a first approximation, it allows us explore the magnitude of the RV bias.
		\end{enumerate}

\begin{figure}[h]
			\centering
			\resizebox{\hsize}{!}{\includegraphics{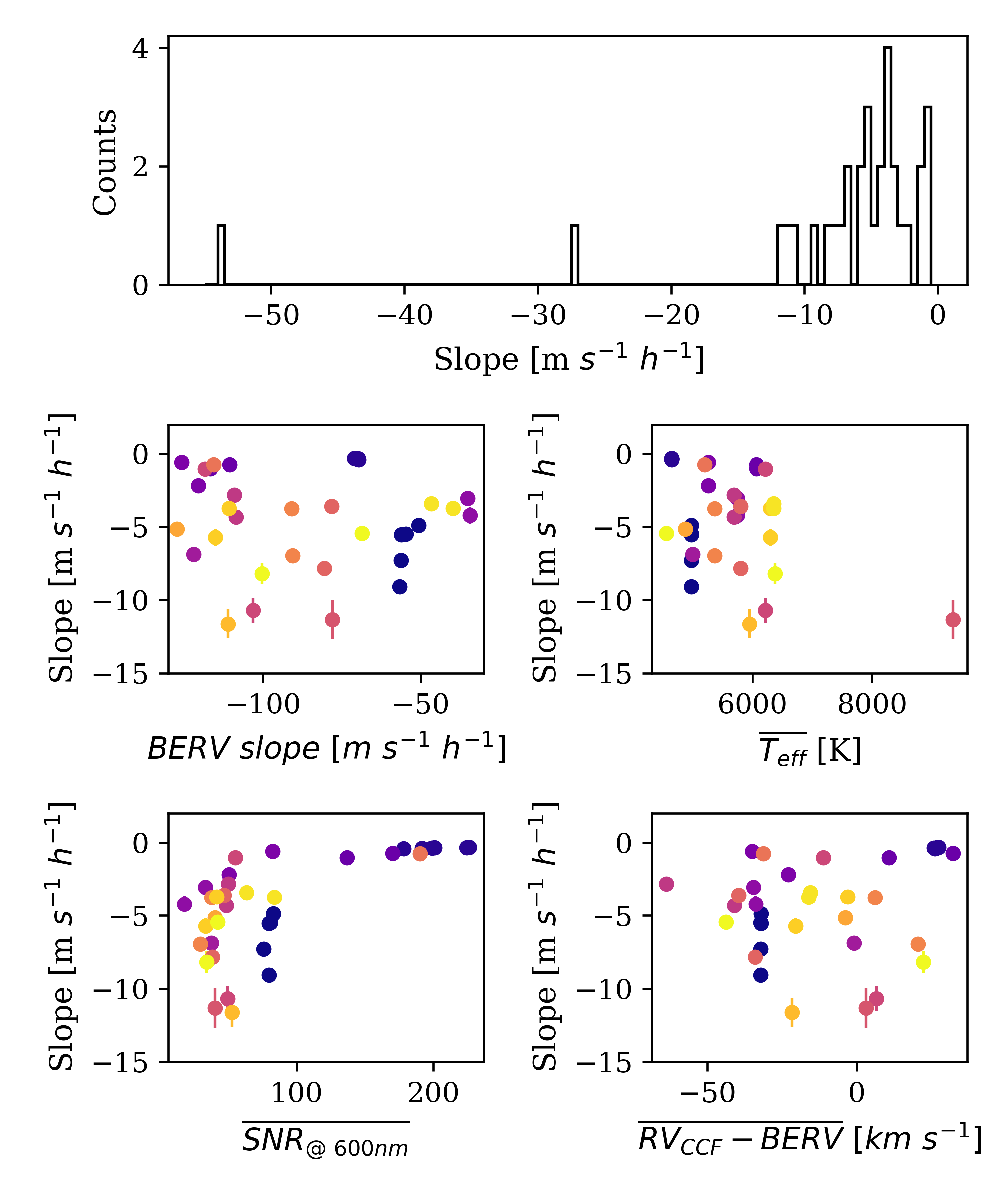}}
			\caption{RV bias for 3785 \espresso{} observations of 20 different stars. The top row presents an histogram of the slopes of a first degree polynomial that was adjusted to the residuals between \sbart{} and CCF RVs. The bottom two rows present a scatter plot with the median value of different metrics: intra-night slope of BERV variation, effective temperature, median signal-to-noise ratio at 600 nm, and median difference between CCF RV and BERV. All observation nights of the same star are shown with a common color. For representation purposes we do not show the two outliers (with a slope greater than -20 \metersecondhour) in the correlation plots.}
			\label{Fig:trend_large_sample_correlations}
		\end{figure}

		Figure \ref{Fig:trend_large_sample_correlations} presents the distribution of the slopes (of the residuals) and their correlation with different metrics, with Appendix \ref{App:slope_of_large_sample} presenting their values. The RV bias that we retrieve always shows a negative slope: independently of the star we find a decrease of RVs over time. This result allows us to discard the hypothesis that the TM algorithm is more sensitive to RV variations introduced by stellar activity, since there is no reason these variations would be in the same direction for all stars.

		For the majority of the targets we find that the absolute value of the residuals RV slope is smaller than 10 \metersecondhour, and the values range from 0.3 \metersecondhour to 53.6 \metersecondhour, with Table \ref{Tab:largesample_slope_list} presenting the individual measurements. It is puzzling why such a large variation is not present on the CCF RVs and does not translate into a change in the line-shape metrics that the CCF provides.
		
		The bottom panels of Figure \ref{Fig:trend_large_sample_correlations} show a scatter plot between the residual slope and different metrics: the intra-night BERV slope; the effective temperature as taken from the \textit{SweetCat} catalogue \citep{sousa_sweet-cat_2024}; the median SNR of the spectra at 600 nm\footnote{We selected \espresso{} order 120, as its median wavelength is the closest to 600 nm.}; and the median difference between BERV and CCF RV. From here we find two noteworthy results: i) The RV bias is not uniform across multiple nights for the same star. An example is \textit{HAT-P-26}, which was observed in two different nights with residual slopes of -27 \metersecondhour{} and -6.9 \metersecondhour; ii) The magnitude of the bias increases as the SNR decreases. We find that for SNR levels under 100 we have a large dispersion of the retrieved RV slopes, whilst for SNR greater than 100, the maximum residual slope found was of -1.02 \metersecondhour;

	\subsection{Wavelength-dependence} \label{Sec:wavelength_depc} 

		One of the core assumptions behind \sbart{} is that a common RV value is enough to describe all differences between the high-SNR stellar model and the spectra. This assumption is in line 	with possible orbiting companions (the signals that we are interested in detecting and characterizing), but not with stellar signals and/or instrumental systematics. 
		
		As a simple test for wavelength dependence of the bias, we compute two different RV time-series: one that only uses the spectral orders that exist in the "blue" detector, and another which only uses information from spectral orders that fall in the "red" detector. We note, however, that the wavelength coverage of the blue and red detectors is not the same for the two spectrographs. 
		
		For this analysis we will use the 2 nights of \HARPS{} observations of HD160691 and 2 of the \espresso{} nights of observations of \epsIndi. We preferred to use the data from \epsIndi{} as it presents a larger flux-SNR than \textit{HD40307}. A subset of the nights was chosen to illustrate the behaviour, as the results were compatible across all nights. 
		
\begin{figure}[h]
			\centering
			\resizebox{\hsize}{!}{\includegraphics{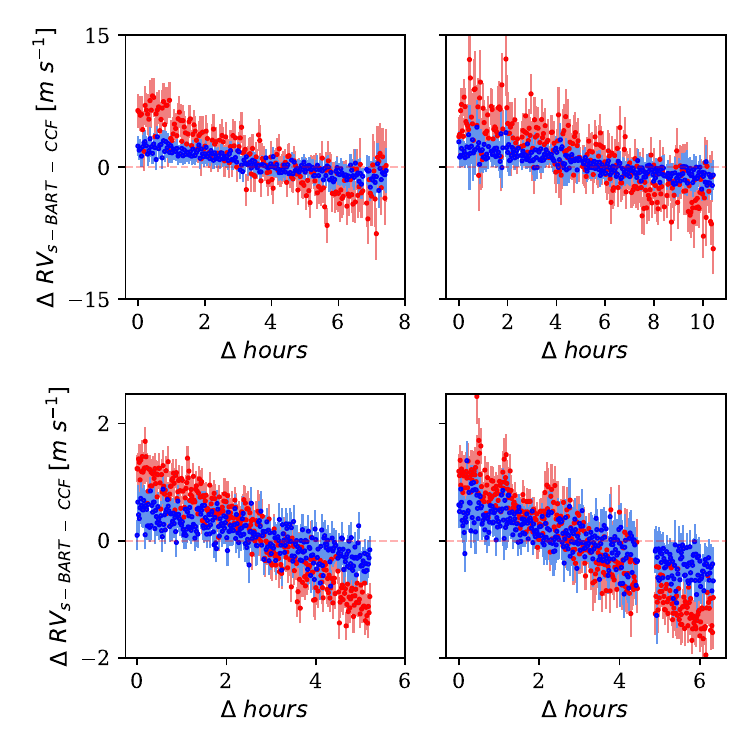}}
			\caption{Residuals between the \sbart{} and CCF RVs for two nights of \HARPS{} observations of HD160691 (top row) and two \espresso{} nights of \epsIndi{} (bottom row). The red and blue points represents the RV time-series extracted when using the corresponding detectors.}
			\label{Fig:trend_mult_stars}
		\end{figure}

		In Figure \ref{Fig:trend_mult_stars} we present the residuals between CCF and \sbart{} RVs for both detectors. We find that in all targets the systematic effect has a larger amplitude in the red detector. The RV time-series from the blue detector still presents a bias, albeit with a smaller amplitude. Unfortunately,  our data does not allow us to draw conclusions from the difference in amplitude between the two spectrographs, as even within a single spectrograph we find large variations (Section \ref{Sect:large_sample_across_targets}).
		
		It is also important to note that the RV time-series that utilizes the full spectral information will reflect the average effect on the two detectors. Albeit not represented here, we also confirmed that this RV drift over time is also seen at the level of the individual spectral orders. From these results we conclude that the bias is present at different wavelengths, showing a small wavelength dependence in amplitude.

\section{A closer look into the BERV-dependency} \label{Sect:tauceti_night_by_night}

	In Section \ref{Sec:trend_introduction} we have shown the presence of a quasi-linear, systematic, bias in RVs collected within a short period of time (i.e., over a few days). The main goal of this section is to understand the impact of relative and absolute BERV variations in the RV bias.
	To measure this dependence, we narrow our analysis to observations of \tauCeti{}, which  was intensively observed during 5 years in the context of \espresso{}'s \textit{GTO}.

	\subsection{Night-by-night analysis of \tauCeti} \label{Sect:tauceti_nightly} 
		
		Selecting a subset of the available nights of  \tauCeti{} observations we compute \sbart{} RVs using different stellar templates:

		\begin{enumerate}
			\item We assume that that each night is fully independent from the others, constructing individual stellar templates for each. The RV time-series obtained from this extraction will denoted as "single-night";
			\item One single stellar template, constructed from all available observations, as used in \citet{figueira_figueiraespressogto_nodate}.
		\end{enumerate}

		As before, to ensure consistency between the different nights we handle telluric features through a common mask. We create a binary telluric mask that rejects wavelength regions with a drop of transmittance greater than 1\%, which is then expanded to account for the maximum yearly BERV variation (which for \tauCeti{} is $\sim$ 29.05 \kilometersecond). This guarantees that the RV extraction is always informed by the same wavelength regions, independently of the position of the telluric features. Similarly to before, we evaluate the magnitude of the bias through a linear fit to the residuals. 

\begin{figure}[h!]
			\centering
			\resizebox{\hsize}{!}{\includegraphics{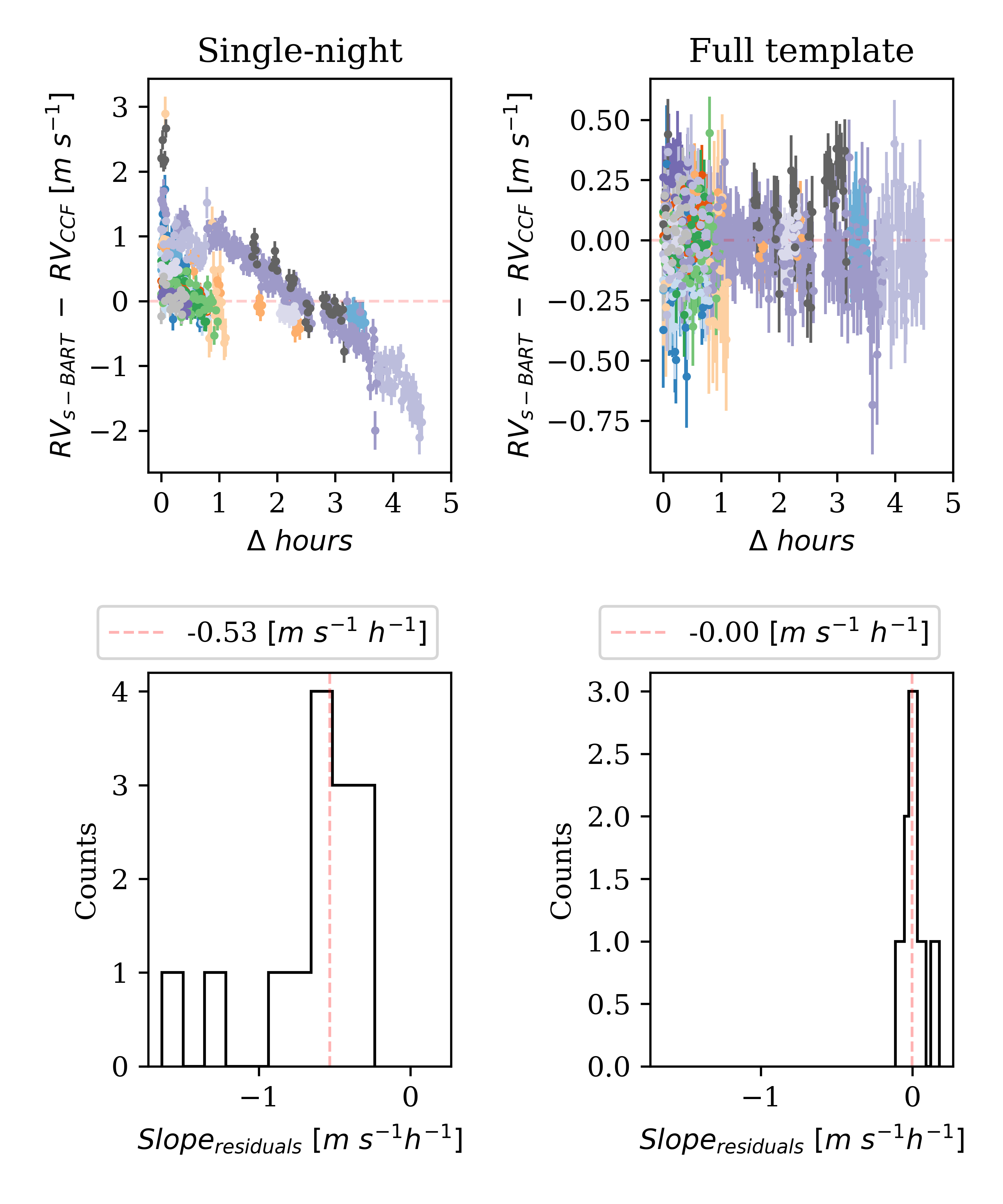}}
			\caption{\textbf{Top:} Residuals between \sbart{} RVs and CCF RVs of \tauCeti{} when using the template constructed from a single night (\textbf{left}) and the one constructed from all available observations (\textbf{right}). Each night is represented by a unique color in the two time-series; \textbf{Bottom:} Histogram of the slopes of the first-degree polynomials adjusted to the residuals of the corresponding column.}  
			\label{Fig:tauceti_single_nights}
		\end{figure}

		Figure \ref{Fig:tauceti_single_nights} shows that the systematic bias is closely linked to the observations that are selected to construct the stellar template. If the template is constructed from observations that are spread over the year we do not find any structure in the residuals (top right panel) nor any meaningful slope within the error bars (median value of $\sim$ 0 \metersecondhour). In the case where the nights are assumed to be independent (left column), each with its own stellar template, there is a linear structure in the residuals, with the maximum slope reaching -1.5 \metersecondhour. This is a clear indication that the origin of this systematic bias is in the construction of the stellar template and the observations that are selected to be part of it.

		Once again we confirm that the linear bias has a negative slope and that it manifests with different amplitudes in different nights. We can also see that even within the data of the same target, we find a difference of 1 \metersecond{} on the residual slope. Furthermore, we also fail to find (Appendix \ref{App:weather_correlation}) any correlation with the meteorological conditions of the individual nights.	

	\subsection{Binning observations based on their BERV} \label{Sect:bervbin}  
		To disentangle a dependence with time from a dependence with BERV (which, within one night would be equivalent) we bin all of our observations based on their BERV. We construct bins of 600 \metersecond, a value close to the expected BERV variation found within a single night of observations. For this, we apply the following prescription:

		\begin{enumerate} 
			\item To avoid having a mixture of red-shifted and blue-shifted telluric features, within a given bin, we remove all observations for which the derivative of the BERV (over multiple nights) is positive. The resulting dataset is highlighted in Figure \ref{fig:TauCetiBERV} and ensures that, within a given bin, the telluric features are all drifting in the same "direction";
			\item We divide our observations into the two \espresso{} datasets -- \textit{ESPRESSO18} and \textit{ESPRESSO19} -- as they will be treated independently during the RV extraction process; This division of \espresso{} data is commonly used and motivated by an instrumental intervention that translated into a change of the spectral profile\footnote{Further details can be consulted in \url{https://www.eso.org/sci/facilities/paranal/instruments/espresso/news.html}}.
			\item We find the minimum BERV of each of the datasets and construct intervals of 600 \metersecond{} starting from it. 
			\item We iterate over all bins and discard those that have less than 20 observations. This ensures that we have a reasonable amount of observations to construct the stellar model and to draw any meaningful conclusion on the presence of a systematic bias.
		\end{enumerate}

\begin{figure}[h]
			\centering
			\resizebox{\hsize}{!}{\includegraphics{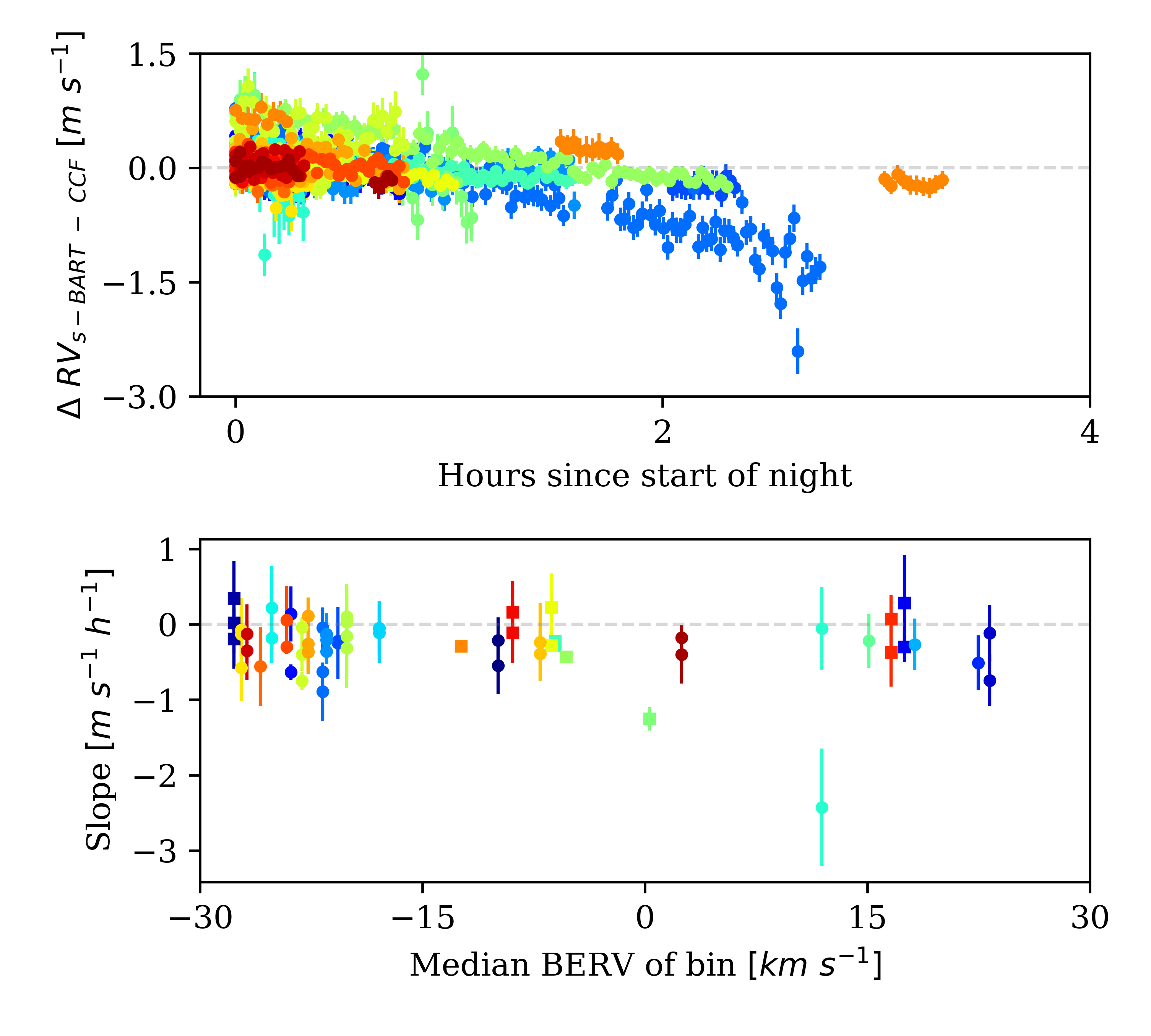}}
			\caption{RV extraction on the \espresso{} dataset of \tauCeti, after binning observations based on their BERV. Each bin is attributed a unique color.\textbf{Top panel:} Residuals between the RV timeseries of \sbart{} and the CCF. To ease visualization we plot the residuals as a function of the time difference between the observation and the first observation of the night it was taken in. \textbf{Bottom panel:} Slope of the first degree polynomial that is fitted to the RV residuals, as a function of the median BERV in the corresponding bin. The squares represent the bins that only contain data for one year, whilst the circles have 2 or more years of observations.}
			\label{fig:TauCetiBERVbin_results}
		\end{figure}

		We are left with 6 \textit{ESPRESSO18} bins and 24 \textit{ESPRESSO19} bins, to which we apply \sbart{}, following the same prescription that has been used throughout the paper. The results from this application are shown in Figure \ref{fig:TauCetiBERVbin_results}, where we show the residuals as a function of hourly difference to the first observation of the corresponding night.  Once again we find a systematic bias within the nights that are included within the bins, with peak-to-peak amplitudes comparable with our previous results. Contrasting previous results, we find that some of the bins present a positive slope, albeit without any correlation with the median BERV of the corresponding bin. Furthermore, we find no obvious difference between the bins that only contain data collected during the same year (squares from the bottom panel) or that contain observations collected over multiple years (circles).

\section{On the origin of the RV bias} \label{Sec:the_hypothesis_of_tellurics} 

	Section \ref{Sect:tauceti_night_by_night} showed that the selection of observations for the construction of the stellar template plays a key role in the presence (or lack) of the RV bias. The templates constructed with the full \tauCeti{} dataset lead to no meaningful RV slopes in individual nights (median value of 0.0 \metersecondhour), whilst the construction of the template from the data of the same night leads to a median slope of -0.53 \metersecondhour{}.	This leads us to think that the origin of this effect comes from the difference between datasets collected within a single night or over time.	
		
	If we focus on the fact that the BERV separation in single-night observations is of the same order of magnitude as the mean pixel size of \espresso{} we can postulate two sources of contamination on the stellar template: micro-telluric features and/or any detector fixed pattern noise. 	We know that in multi-night observations micro-tellurics can -- independently of the RV-extraction method used -- impact RV measurements at the 10 \centimetersecond{} level \citep{cunhaImpactMicrotelluricLines2014} and bias our RV extraction. If we think about their impact on the stellar template, we see that they will be -- assuming a large BERV span and an high number of observations -- averaged out of the stellar template. They are, by definition, shallow features and when combining the multiple observations to construct the template, they will be averaged out of it. In the same vein, any instrumental systematic, or correlated noise, that is coherent over a timescale of a few hours will also be averaged out of the stellar template in the case of multi-night observations.
	
	In single-night observations, those assumptions might no longer hold. Since the BERV separation within our dataset is small, any contamination on the detector referential, either from tellurics or from artifacts on the detector, will be approximately in the same position and overlap over time. The BERV variation during one night is smaller than the RV coverage of one pixel and than the Full Width at Half Maximum (FWHM) of spectral lines. Thus, when stacking our observations (through a weighted mean, or equivalent) they will not be averaged out. In the specific case of micro-telluric lines, they will be appear in the template slightly broader and shallower than in any individual observation. In the case of fixed pattern noise on the detector, the stellar template will contain the mean effect across the night, also blurred by the BERV variation. Afterwards, in the RV extraction process, those features and any other instrumental contamination could be driving the alignment between spectra and template as, during the night, they will be shifting in the same "direction" in the spectra.

	This contamination from micro-tellurics and correlated noise would provide an explanation of the quasi-linear shape of the bias and the fact that all observations present a negative slope (Section \ref{Sect:large_sample_across_targets}), as the light moves in the detector in the same direction due to the negative slope of the BERV variation over the night. It would also explain the fact that the bias is also present in \serval{} and \arve{} RV measurements, as they also utilize a stellar template constructed from stacking multiple observations. It is also possible to justify the lack of effect on CCF RVs through a combination of two different factors: i) The CCF method uses a fixed mask through time, leading to a model (the CCF mask) that is not contaminated; ii) the CCF mask only includes a small subset of the spectral lines (and regions) that are present in the stellar template. 
	
	Overall, this hypothesis seems to be compatible with the results that we report in this work:

	\begin{itemize}
		\item In Section \ref{Sec:wavelength_depc}, with a chromatic analysis, we report a greater effect in the red wavelengths, where telluric features are more common.
		\item  In Section \ref{Sect:bervbin} we binned \tauCeti{} observations based on their BERV and found the RV bias, even when combining observations of multiple years. This gives weight to the hypothesis that the RV bias is connected with the physical position of the spectral  lines on the detector. We also saw that increasing the number of observations that are used to construct the stellar model (and, consequently, increase the BERV span) leads to a decrease of the RV slope, until we achieve converge towards a slope of zero, as one would find in multi-night datasets. 
		\item We find that a time-separation between single-night spectra and respective template lead to a decrease of the effect (Appendix \ref{Sec:berv_separation_impact}). This is still compatible with our two sources of contamination, as the BERV separation will: i) place the micro-tellurics of the template further way from the ones in the spectra; ii) lead to a time-evolution of the correlated noise sources, diluting its impact on the RV extraction as it no longer matches the effect imprinted on the first night.
	\end{itemize}

	In Section \ref{Sec:tauceti_telluric_threshold} and \ref{Sect:berv_span_impact} we further explore the hypothesis of telluric contamination in the data, through changes in the telluric masking and the BERV of the observations.
		
	\subsection{Impact of the telluric threshold} \label{Sec:tauceti_telluric_threshold} 

		Here we study the effects of telluric contamination by changing the  threshold to reject telluric-contaminated regions. A decrease in this threshold will result in a higher number of (deeper) features bypassing our rejection criteria and, consequently, increase this contamination of our observations. Thus, if the RV trend is introduced by leftover tellurics in the stellar template, as described in Section \ref{Sec:the_hypothesis_of_tellurics}, we expect the amplitude to increase as our rejection criterion gets more relaxed.

		To evaluate this hypothesis, we restrict our nights even further: from the subset of nights presented in Table \ref{Tab:tauceti_nights_summary} we select those for which the last observation was taken more than 3 hours after the first one. This choice leaves us with three nights of data and its selection was motivated by visualization purposes: the systematic bias is clearer over longer time-scales. Then, we extract RVs in the following fashion:

		\begin{enumerate}
			\item We construct a new binary mask using a given threshold (\textit{$Tr_{i}$}) for the telluric contamination; It is important to note in here that \sbart{} uses this threshold as a percentage of the continuum level. If \textit{$Tr_{i}$} = 1\%, it rejects every wavelength in which the continuum varies more than 1\%. This means that higher values of \textit{$Tr_{i}$} reject less telluric features than lower ones. For further details on the construction of the telluric mask we refer back to \citet{silvaNovelFrameworkSemiBayesian2022}.
			\item For each night in our restricted set (i.e, the 3 nights selected based on having a time-span grater than 3 hours) we use the telluric template from 1) and construct a new stellar model. Then, we apply the usual \sbart{} RV extraction to get a new RV time-series.
		\end{enumerate}

\begin{figure}[h]
			\centering
			\resizebox{\hsize}{!}{\includegraphics{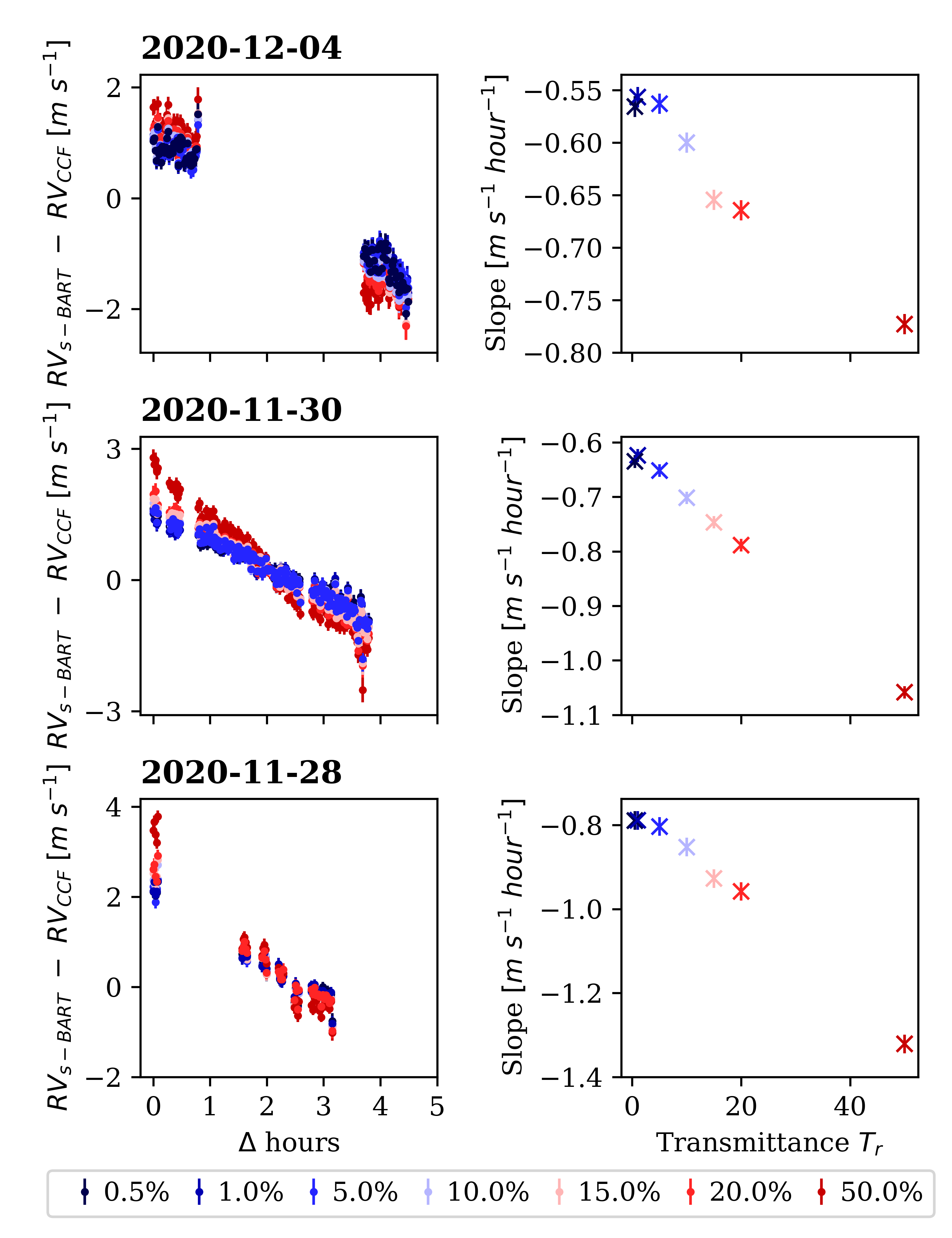}}
			\caption{Impact on the RV trend from template matching after creating a telluric mask with different thresholds for the rejection of telluric-contaminated regions. The thresholds were applied to the same transmittance spectrum, generated through \textit{Telfit} \citep{gulliksonCorrectingTelluricAbsorption2014}. Each row represents different nights of \tauCeti, chosen from within our subset specified in Table \ref{Tab:tauceti_nights_summary}. \textbf{Left:} Residuals to the CCF RVs, color-coded based on the transmittance threshold; \textbf{Right:} Slope of a first-degree polynomial adjusted to the residuals as a function of the transmittance threshold.}
			\label{Fig:tauceti_transmit_change}
		\end{figure}

		We repeat this process for a list of seven different thresholds -- with \textit{$Tr_{i}$} taking the values of $\{0.5, 1, 5, 10, 15, 20, 50\}$ \% -- and compare the RV time-series with the one that is extracted from the CCF, as shown in Figure \ref{Fig:tauceti_transmit_change}.

		From figure \ref{Fig:tauceti_transmit_change} we find that a relaxation in the masking of the telluric features (increasing the threshold) leads to an increased  effect in our RVs, which lead to a greater difference to the ones from the CCF. This is true for all cases, except for a telluric rejection of 0.5\%, where the slope is slightly larger than for a threshold of 1\%. With such a small threshold (half of \sbart{} default value of 1\%) we have a massive rejection of wavelength regions, leading to a noisier time-series that still leads to a slope that is compatible at the 1-$\sigma$ level.

		If we now focus on the difference between the results obtained with a telluric mask with \textit{$Tr_{i}$} of 50\% and the one with \textit{$Tr_{i}$} of 1 \% we find an interesting result. On the two nights from November we find a decrease in the RV slope of $\sim$ 40\%, whilst on the night from the December the telluric threshold variation only translates to a decrease of $\sim$ 27\%. Furthermore, the night from December also presents the smaller peak-to-peak variation, the larger time-span between start and end of observations, and the  larger gap in the middle of the observations. This results seem to be compatible with what one would expect from the presence of telluric features: The wide gap in the data from December eases the "averaging-out" of the noise sources from the template. Thus, the extraction is not as sensitive (in comparison with the other nights) to their presence.
		
		Independently of the threshold that we use, we still find the quasi-linear trend between the two RV time-series.  In Appendix \ref{App:TCtellcorr} we present an overview of the effect that is introduced on data that was corrected from telluric features, showing results compatible, at the 1-$\sigma$ level to those that are present in here.
	
	\subsection{Impact of BERV span on recovered RVs} \label{Sect:berv_span_impact} 
		
		Throughout this Section we have been evaluating a possible contamination of telluric features and correlated noise sources on the stellar template. If this is indeed the case, we should find that as we increase the temporal span of our dataset (i.e, the observations that are used to construct the stellar template) the bias should decrease. For this analysis, we go back to our sub-set of pre-selected nights (Table \ref{Tab:tauceti_nights_summary}) and we sequentially add more observations to the stellar template in the following fashion:

		\begin{enumerate}
			\item We select the $i^{\rm th}$ night and construct a stellar template using all observations taken until the end of that night (i.e., all observations since the start of the first night);
			\item We extract RVs from all observations collected in the night of the 28th of December, 2020, whilst using the newly created stellar template. During this process we use the telluric mask with a transmittance cutoff of 1\%.
		\end{enumerate}
		
\begin{figure}[H]
			\centering
			\resizebox{\hsize}{!}{\includegraphics{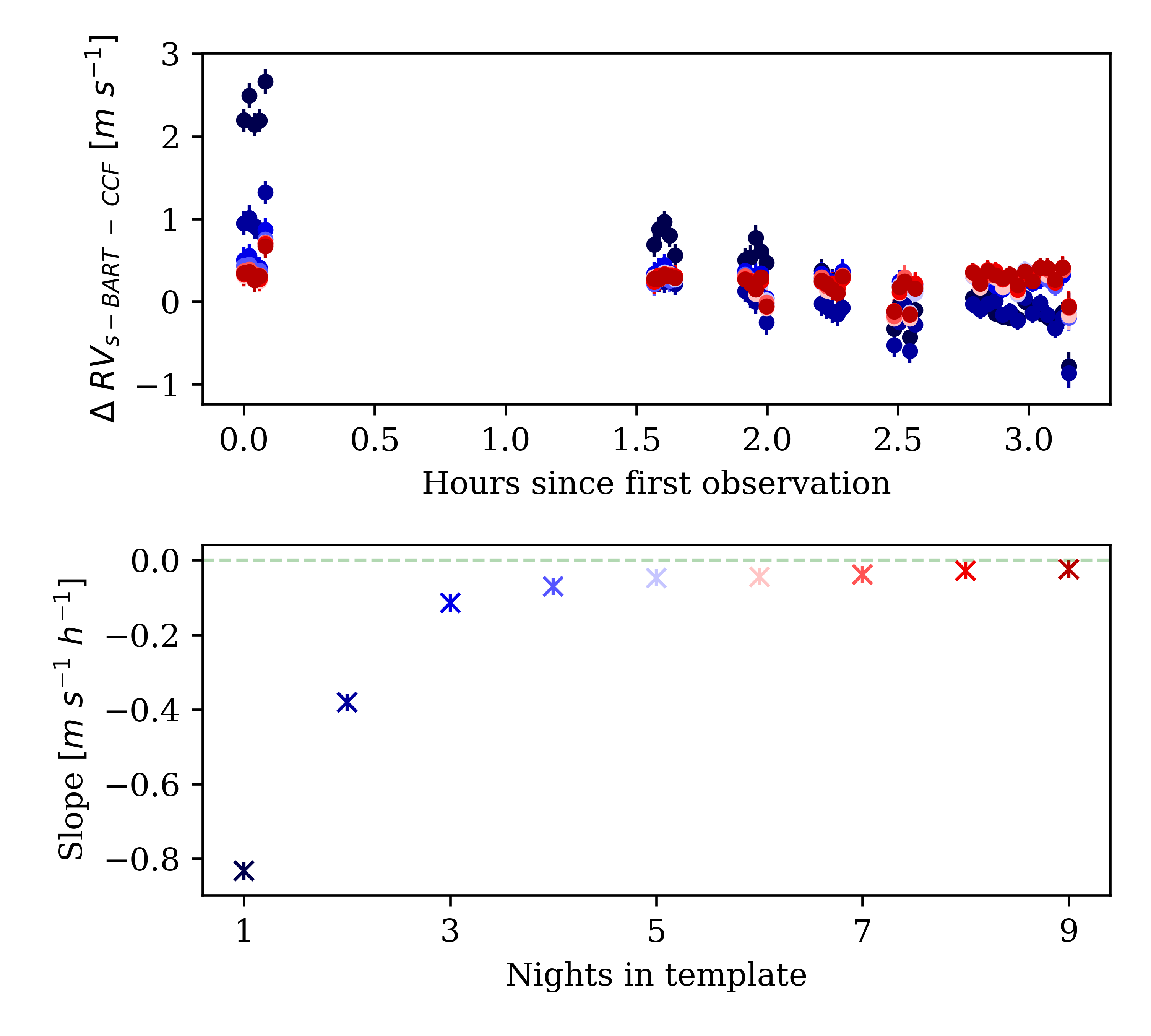}}
			\caption{\textbf{Top panel:} Comparison of different RV time-series of \tauCeti, extracted with stellar templates that sequentially utilise one more night of observations than the previous; \textbf{Bottom panel:} Slope of a first degree polynomial adjusted to the residuals as a function of the number of nights in the stellar template.}
			\label{Fig:tauceti_single_nights_roll_template}
		\end{figure}

		In Figure \ref{Fig:tauceti_single_nights_roll_template} we present the results from this comparison, where it is clear that the addition of new observations into the template leads to a significant decrease in the RV systematic bias. In there, we see that the slope of the residuals quickly converges to a near-zero value as the stellar template includes observations from more than 5 nights.

		It is also important to note that the \tauCeti{} observations present some of the lowest trends in our datasets, which might mean that the results from Figure \ref{Fig:tauceti_single_nights_roll_template} are a best-case scenario
		
	\subsection{A summary of the RV bias}

		We found, in Section \ref{Sec:tauceti_telluric_threshold}, that a less conservative telluric mask leads to an increase in the effect. However, even with the most conservative threshold that we use (masking tellurics deeper than 1\%) we are not able to fully reduce the RV slope to 0 \metersecondhour. However, it is also important to keep in mind that this test only probes the effect on the lines that are present on the transmittance model. Our results can be interpreted in different ways: i) the models that we use to construct the telluric binary mask are missing lines; ii) telluric contamination could introduce a similar effect to what we see, but it is not the root cause; or iii) The effect that we measure is a combination of different sources of contamination of the stellar template. The first hypothesis is difficult to support based on our data, as both the telluric masking and telluric correction (described in Appendix \ref{App:TCtellcorr}) show compatible results. The increase of the absolute RV slope as we relax the telluric threshold does not allow us to confidently claim the telluric features to be the root source of this issue. If we consider a given fixed-pattern noise artifact that exists on the referential of the detector, it will introduce a flux contamination similarly to the one introduced by telluric features. During the course of one night, the physical drift of stellar lines on the detector leads to them being imprinted on the data, similarly to how telluric features are overlaped with the stellar spectra. Under this hypothesis, a given percentual masking of telluric features minimizes the RV residuals introduced by them, still keeping the contamination from the fixed-pattern artifact. A relaxation of the telluric masking translates to a magnified effect since the unmasked tellurics would introduce a new systematic RV slope to the pre-existing one from a fixed-patttern artifact. Based on our current results and understanding, we postulate that this RV systematic trend is introduced by a combination of micro-telluric features and other, unidentified, artifacts on the detector that are coherent over timescales of a few hours.	
		
		The fact that the bias decreases as we increase the time between the data and template (Figure \ref{fig:TauCeti_year_of_data}) is fully expected, since we are left with a larger separation between the systematics of the template and the data, breaking the coherence of the flux artifacts. Furthermore, a combination of observations from different nights (as shown in Figure \ref{Fig:tauceti_single_nights_roll_template}) leads to a decrease of the RV trend, which is compatible with both sources of contamination. The larger BERV span results in a larger separation between the telluric lines across all observations, diluting their presence in the stellar template. Furthermore, the temporal separation can reduce the coherence of the fixed pattern systematics present on the spectra. Both of those effects contribute to a dilution of the contamination of the stellar template and, consequently, a decrease in the RV trend that we measure. In Appendix \ref{App:othercauses} we discuss other possible sources of noise on single-night observations.

\section{Discussion} \label{Sec:discussion} 

	In this paper we present and explore a previously unidentified  systematic RV bias on template-based RVs that is not found on CCF radial velocities. This effect is present for several stars observed with different spectrographs (see Section \ref{Sect:large_sample_across_targets}) with residual RV (of TM compared with CCF) slopes in the interval [-0.3, -52] \metersecondhour. Furthermore, we find that the amplitude of the RV residuals is not consistent across different nights of observations of the same target. The fact that this trend is found across different spectrographs, as well in different pipelines, hints towards a systematic effect at play that is biasing our RV extraction. More worrisome is that not only do we have to consider the impact on intra-night observations, but also understand how it translates to RV extraction in multi-night observations. Fortunately, as we have shown in Section \ref{Sect:tauceti_nightly}, this bias is either not present under such conditions, or exists at an amplitude that sits under the noise floor of our instruments.

	\subsection{Impact on different scientific cases} 

		In this paper we have explored a systematic RV signal that affects our intra-night measurements with state-of-the-art high-resolution optical spectrographs, such as \espresso{} and \HARPS{}. This contamination will have an effect on the RV measurements that we obtain for different scientific purposes. Below we discuss the impact on three of those.

		\paragraph{\textit{Detection and characterization of exoplanets:}} The determination of precise RVs plays a pivotal role in the detection and characterization of exoplanets \citep[e.g.,][]{castro-gonzalez_unusually_2023, mascareno_two_2023,gonzalez_hernandez_sub-earth-mass_2024, passegger_compact_2024}. Under such scientific goals, stellar spectra is usually acquired within a large window of BERVs and, consequently, template-matching RVs are not affected by this bias. The effect that we report is averaged out from the stellar template (given a certain number of observations spread over different nights) to a level that cannot be measured on RV time-series with current spectrographs. However, at shorter timespans of observations, this time-correlated effect could be interpreted as a planetary signal. Fully understanding this spurious effect and developing a robust enough model will be critical to exploit the stellar spectra from the next generations of spectrographs and to possibly reduce the current noise-floor of our instruments.
		
		\paragraph{\textit{Impact on transit and atmospheric characterization:}} Transit observations have become central to modern astrophysics, providing key insights into the architecture of planetary systems \citep{winn_hot_2010, albrecht_stellar_2022,cristo_espresso_2024} and enabling the study of exoplanet atmospheres via transmission spectroscopy \citep{casasayas-barris_atmospheric_2019, ehrenreich_nightside_2020,azevedo_silva_detection_2022}. A crucial method for determining the relative orientation between a star's rotation axis and a planet's orbital plane is the Rossiter-McLaughlin (RM) effect \citep{holt_spectroscopic_1893, rossiter_detection_1924,mclaughlin_results_1924}. \citet{boue_new_2013} demonstrated that the RV signal from the RM effect is sensitive to the method used for RV extraction. Therefore, understanding the application of template-based RVs in this context is essential, particularly in determining how any spurious slope could affect the interpretation of the RM curve. In transmission spectroscopy, the standard approach for constructing absorption spectra involves creating a template (or "master-out") from out-of-transit observations, which is then compared to in-transit spectra. However, it is possible that a spurious contamination of the master-out (and associated RV measurements) may contaminate in-transit absorption spectra, potentially introducing biases in the determination of planetary properties or leading to false species detections, especially when both the behind-the-planet stellar regions and the planet have similar velocities.
		
		\paragraph{\textit{Asteroseismology impact:}} Prior to the mid-2000s, RV observations constituted the primary method for measuring convection-driven, solar-like oscillations in stars other than the Sun, leading to detections for a few select stars \citep[see, e.g.,][]{kjeldsen_amplitude_2008}. The advent of space photometry, with missions such as CoRoT \citep{auvergne_corot_2009}, Kepler/K2 \citep{koch_kepler_2010,howell_k2_2014}, and TESS \citep{ricker_transiting_2014}, has since revolutionized the field, enabling the asteroseismic analysis of hundreds of thousands of stars \citep{campante_asteroseismic_2016,hon_quick_2021}. Despite these advancements, there remained a notable detection void in cool main-sequence stars (i.e., cooler than $\sim$ 5000 K). Owing to their low luminosities, K dwarfs display extremely small oscillation amplitudes (below the 10 cm/s level or, equivalently, a few parts-per-million), which are thus beyond the reach of current space-based photometric missions. A viable alternative is offered by the much higher SNR in RV observations over the typical oscillation timescales, which has been driving a resurgence of observing campaigns on K dwarfs using state-of-the-art spectrographs, such as ESPRESSO \citep{pepeESPRESSOVLTOnsky2021} and KPF \citep{gibson_keck_2020}, mounted on large-aperture telescopes \citep{campante_expanding_2024,hon_asteroseismology_2024,li_k-dwarf_2025}. The short-cadence (< $\sim$ 60 s) and continuous observations needed to properly sample these oscillations make it, however, impractical to secure more than a handful of consecutive nights at the telescope, which necessarily translates into short observational baselines. It is clear throughout this paper that this systematic bias in the RV extraction presents a critical challenge in the utilization of template-based RV time-series in the context of asteroseismology studies.
	
	\subsection{Possible avenues to decrease the impact} \label{Sec:impact_decrease} 
		
		Throughout the paper we explored different avenues to mitigate the impact of this contamination of the stellar template. In Section \ref{Sect:tauceti_nightly} we have seen that the construction of a template from a large number of observations -- spread through multiple years -- is one possible avenue to mitigate this systematic effect. This was also seen in Section \ref{Sect:berv_span_impact} where the addition of new nights of observations leads to a gradual decrease of the RV slope. In the context of exoplanetary detection and characterization -- of exoplanets with orbital periods with timescales greater than a few days -- such observations will be collected over time, leading to a natural mitigation of this systematic bias. However, this is not the case for all observational campaigns, as some consist of a short number of nights with high cadence (e.g., asteroseismology, transit and atmospheric studies, characterization of ultra-short period exoplanets). It is thus of paramount importance that, for such science cases, observations are collected over a larger window of BERVs, so that the construction of data-driven stellar models is not impacted by this effect.
		
		The option of using observations collected after a certain time has passed (as shown in Appendix \ref{Sec:berv_separation_impact}) also seems to lead to satisfactory results on the higher-SNR stars. However, it also seems to be accompanied with an increased RV scatter. It was not clear if this excess of variability could also be a byproduct of the root source of the systematic bias. The interpretation of the lower SNR case is made more difficult by the shorter coverage of HD40307 (5 nights, in comparison with the previous 6) and the overlap in BERV that we have between observations (see Appendix \ref{App:dset_overview}). Albeit we find a similar behaviour to the one from the high-SNR case, it comes with a delay of 1 day. Thus, we are not able to properly interpret the lower SNR case of the asteroseismology datasets. 

		In the case where all options above are not valid, it would also be possible to use a stellar model constructed with data from a different star. If we find a star that has similar properties to the one in which we are interested in, we can construct the stellar template with its data and then apply it to our original star \citep[e.g., as described in][]{artigau_line-by-line_2022}. The core concept behind template-matching algorithms is that the stellar model should be a perfect match to the observations of our star. By using a model from a different star we will be introducing differences between model and data, that would introduce extra variability on the results.

\section{Conclusions} 

	In this paper we presented a systematic RV effect that is contaminating template-matching (TM) and line-by-line (LBL) RVs in the specific case where all observations fall within a short time-window (i.e., within a few days). Not only do we find this effect in different template-based pipelines (\sbart{}, \serval{}, and \arve), but also on different stars and spectrographs, with slopes that vary, in our sample, in the interval [-0.3, -50] \metersecondhour. This systematic effect presents a significant challenge for the usage of template-matching RVs for science cases that focus on extensive, intra-night, observations, e.g., asteroseismology analysis, transit and atmospheric characterization.
	
	From \espresso{}  and \HARPS{} observations we find that the contamination can be found on each detector, with it being systematically larger on the red detector. We also find that this effect either doesn't come into play for multi-night datasets or it exists under the current noise-floor of our instruments and analysis techniques.	We managed to pin-point the source of this systematic bias to the construction of the stellar template. In particular, to the selection of observations from which it is constructed. We postulate that a combination of leftover micro-telluric features and correlated noise in the stellar template leads to the systematic bias of the RVs over the night. We found that it can be mitigated by either using a large temporal separation between template and data (in the case where the template is also constructed from one single night) or by using a large number of observations collected over a period of multiple months.

\begin{acknowledgements}
	We thank the helpful comments of Andrew Collier Cameron as the referee of this paper. This work was funded by the European Union (ERC, FIERCE, 101052347). Views and opinions expressed are however those of the author(s) only and do not necessarily reflect those of the European Union or the European Research Council. Neither the European Union nor the granting authority can be held responsible for them. This work was also supported by FCT - Fundação para a Ciência e a Tecnologia through national funds by these grants: UIDB/04434/2020 DOI: 10.54499/UIDB/04434/2020, UIDP/04434/2020 DOI: 10.54499/UIDP/04434/2020, PTDC/FIS-AST/4862/2020, UID/04434/2025. TLC is supported by Funda\c c\~ao para a Ci\^encia e a Tecnologia (FCT) in the form of a work contract (\href{https://doi.org/10.54499/2023.08117.CEECIND/CP2839/CT0004}{2023.08117.CEECIND/CP2839/CT0004}). J. H. C. M. acknowledges further support from national funds through the FCT DarkMAGE project (grant ID: PTDC/FIS-AST/4862/2020) and from  the project e-CHEOPS (PEA: 4000142255), funded by ESA/PRODEX. JIGH acknowledge financial support from the Spanish Ministry of Science, Innovation and Universities (MICIU) projects PID2020-117493GB-I00 and PID2023-149982NB-I00. ÉA is supported by the Trottier Family Foundation through the Trottier Institute for Research on Exoplanets (IREx). ODSD acknowledges support from e-CHEOPS: PEA No 4000142255. The INAF authors acknowledge financial support of the Italian Ministry of Education, University, and Research	with PRIN 201278X4FL and the "Progetti Premiali" funding 	scheme. X.De acknowledges funding from the French ANR under contract number ANR-24-CE49-3397 (ORVET). This work is supported by the French National Research Agency in the framework of the Investissements d’Avenir program (ANR-15-IDEX-02), through the funding of the “Origin of Life" project of the Grenoble-Alpes University. KA acknowledges support from the Swiss National Science Foundation (SNSF) under the Postdoc Mobility grant P500PT\_230225. J.L.-B. is funded by the Spanish Ministry of SCience, Innovation and Universities (MCIN/AEI/10.13039/501100011033) through grants PID2019-107061GB-C61, PID2023-150468NB-I00 and CNS2023-144309. We acknowledge financial support from the Agencia Estatal de Investigaci\'on of the Ministerio de Ciencia e Innovaci\'on MCIN/AEI/10.13039/501100011033 and the ERDF “A way of making Europe” through project PID2021-125627OB-C32, and from the Centre of Excellence “Severo Ochoa” award to the Instituto de Astrofisica de Canarias. X.Du acknowledges the support from the European Research Council (ERC) under the European Union’s Horizon 2020 research and innovation programme (grant agreement SCORE No 851555) and from the Swiss National Science Foundation under the grant SPECTRE (No 200021\_215200). This work has been carried out within the framework of the NCCR PlanetS supported by the Swiss National Science Foundation under grants 51NF40\_182901 and 51NF40\_205606. FPE would like to acknowledge the Swiss National Science Foundation (SNSF) for supporting research with ESPRESSO through the SNSF grants nr. 140649, 152721, 166227, 184618 and 215190. The ESPRESSO Instrument Project was partially funded through SNSF’s FLARE Programme for large infrastructures. This work was financed by Portuguese funds through FCT (Funda\c c\~ao	para a Ci\^encia e a Tecnologia) in the framework of the project 2022.04048.PTDC (Phi in the Sky, DOI 10.54499/2022.04048.PTDC). CJM also acknowledges FCT and POCH/FSE (EC) support through Investigador FCT Contract 2021.01214.CEECIND/CP1658/CT0001 (DOI 10.54499/2021.01214.CEECIND/CP1658/CT0001). ASM acknowledges financial support from the Spanish Ministry of Science,
	Innovation and Universities (MICIU) projects PID2020-117493GB-I00 and 	PID2023-149982NB-I00. ARCS acknowledges support from  FCT fellowship 2021.07856.BD;  This work has received support in the framework of the National Centre of Competence in Research PlanetS supported by the Swiss National Science Foundation under grant 51NF4\_205606. The authors acknowledge financial contribution from the European Union - Next Generation EU RRF M4C2 1.1 PRIN MUR 2022 project 2022CERJ49 (ESPLORA).
\end{acknowledgements}

\bibliographystyle{aa}
\bibliography{main}

\begin{appendix}

	\section{BERV coverage of the sample of stars}
		\begin{figure}[H]
			\centering
			\resizebox{\hsize}{!}{\includegraphics{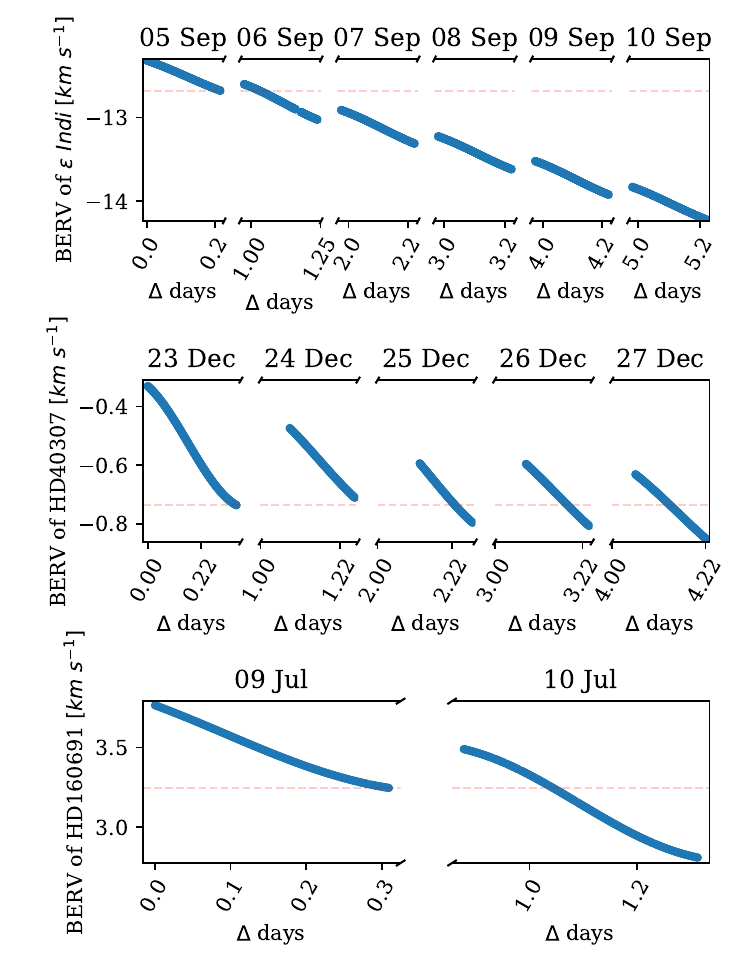}}
			\caption{BERV values for the different nighs of the asteroseismology stars as a function of the temporal distance to the first observation of that star. \textbf{Top and middle row:} \espresso{} observations of  \epsIndi{} and \textit{HD40307}, respectively. \textbf{Bottom row:} \HARPS{} observations of \textit{HD160691}. The dashed red line highlights the BERV value of the last observation taken in the first night.}
			\label{fig:astero_berv_over_night_app}
		\end{figure}
		
		\begin{figure}[H]
			\centering
			\resizebox{\hsize}{!}{\includegraphics{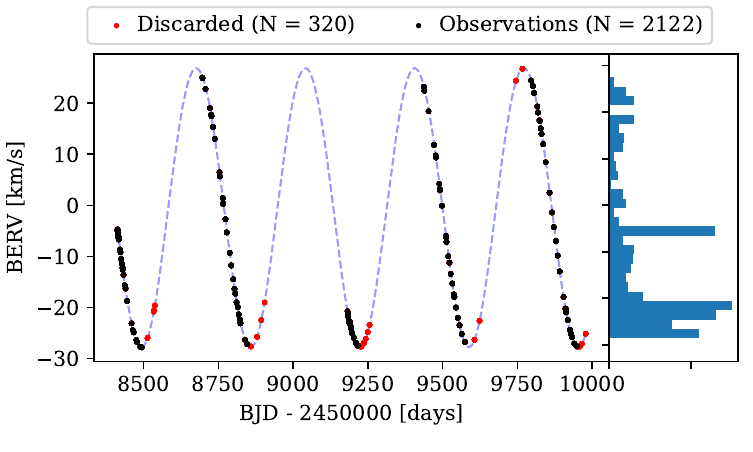}}
			\caption{BERVs of the \espresso{} observations of \tauCeti{} (left) and their distribution (right) in bins of 1 km/s. The black points (2122 in total) represent the observations that we keep for our analysis of Section \ref{Sect:bervbin}, whilst the red points were discarded. The dashed blue line represents the BERV curve of \tauCeti{}.}
			\label{fig:TauCetiBERV}
		\end{figure}
						
		\begin{table}[h]
			\caption{Summary of the data from the three stars (\textit{HD40307}, \epsIndi{}, and \textit{HD160691}) that were selected for our analysis.}
			\label{Tab:astero_snr_over_night_app}
			\centering
			\begin{tabular}{c|ccc}\hline \hline
				Star      & Night &  Median $SNR_{@ 600 nm}$ \\ \hline
				\multirow{ 2}{*}{HD160691 (\HARPS)} 	&  2004-06-09     & 120.15 \\
				 										&   2004-06-10    &  99.18 \\ \hline
				\multirow{ 5}{*}{HD40307 (\espresso)}   & 2018-12-23 & 78.56  \\
													    & 2018-12-24 & 75.62  \\
													    & 2018-12-25 & 76.55  \\
													    & 2018-12-26 & 75.52  \\
													    & 2018-12-27 & 71.96  \\ \hline
				\multirow{ 6}{*}{\epsIndi{} (\espresso)}  & 2022-09-05 & 231.35 \\
														  & 2022-09-06 & 195.03 \\
														  & 2022-09-07 & 229.35 \\
														  & 2022-09-08 & 195.12 \\
														  & 2022-09-09 & 180.74 \\
														  & 2022-09-10 & 201.96 \\ \hline
			\end{tabular}
			\tablefoot{For the median SNR calculation we use the flux SNR for spectral order 57 in \HARPS{} and 120 for \espresso.}
		\end{table}

		\begin{table}[h!]
			\caption{Summary of the observation nights that were used in this analysis, containing the night (left), the number of observations (middle) and the BERV of the first observation (right).}
			\label{Tab:tauceti_nights_summary}
			\centering
			\begin{tabular}{ccc}\hline \hline
				Night      & N observations & BERV [\kilometersecond] \\ \hline
				2020-11-28 & 39             & -20.97                  \\
				2020-11-29 & 35             & -21.54                  \\
				2020-11-30 & 155            & -21.47                  \\
				2020-12-03 & 20             & -22.71                  \\
				2020-12-04 & 80             & -22.68                  \\
				2020-12-05 & 40             & -23.13                  \\
				2020-12-09 & 40             & -23.84                  \\
				2020-12-10 & 40             & -24.1                   \\
				2021-01-11 & 40             & -27.62                  \\ \hline
			\end{tabular}
		\end{table}

		\begin{figure}[H]
			\centering
			\resizebox{\hsize}{!}{\includegraphics{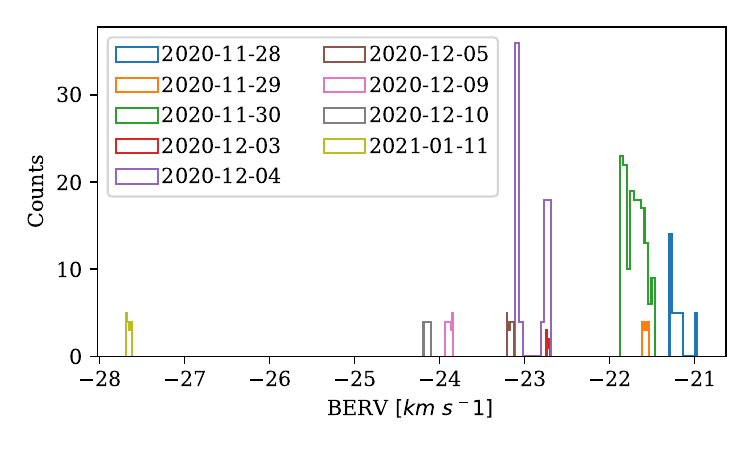}}
			\caption{BERV coverage of the subset of observing nights of \tauCeti{} that we selected to explore the systematic RV bias.}
			\label{Fig:tauceti_single_nights_bervdist}
		\end{figure}

		\begin{figure}[H]
			\centering
			\resizebox{\hsize}{!}{\includegraphics{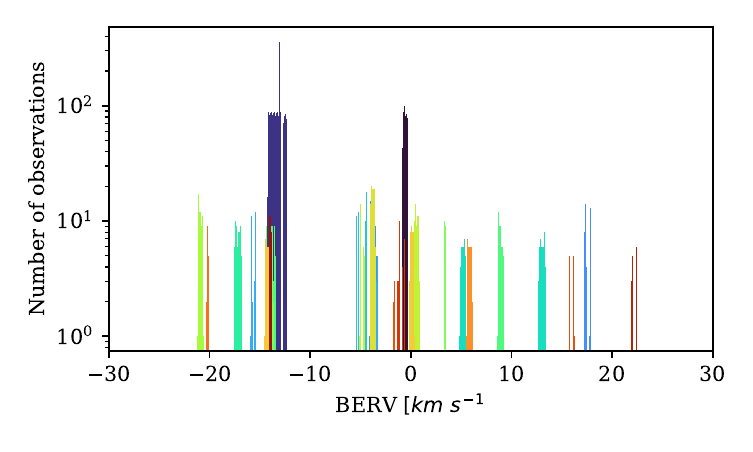}}
			\caption{Histogram of the BERV distribution of all stars in our \espresso-WG2 sample. Multiple nights observations of the same star are shown with the same color.}
			\label{fig:largesample_EBRVdist}
		\end{figure}

	\section{Searching for correlations with line-shape evolution} \label{Sec:correlation_search_large_sample} 
\begin{figure*}
		\centering
		\includegraphics[width=17cm]{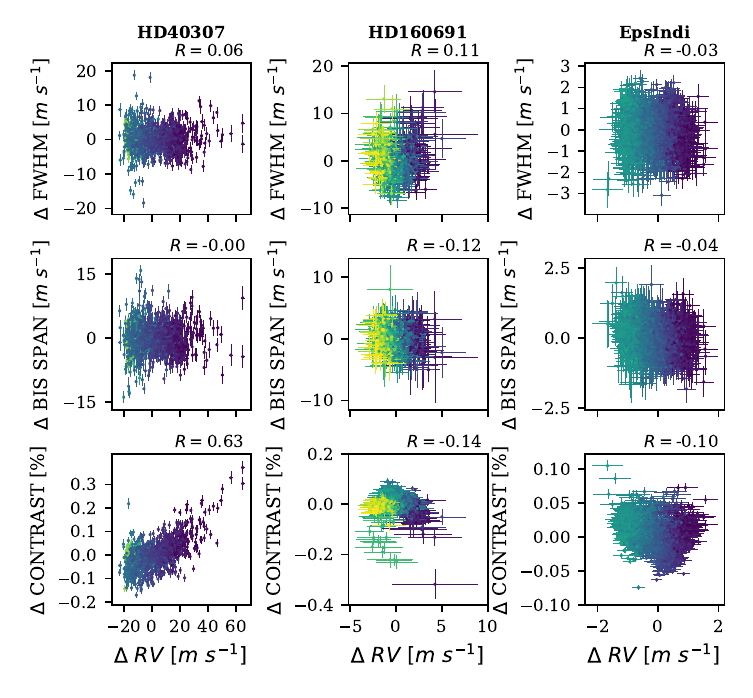}
		\caption{Correlation between the \sbart{} and CCF RVs for HD40307 (\espresso, left column), HD160691 (\HARPS, middle column), and \epsIndi{} (\espresso, right column). In each row we plot the values from the activity indicators from the CCF: FWHM (top row), BIS SPAN (middle row), and Contrast (bottom row). As we are including more than one night, we remove the median value of each indicator, so that they are centered at zero. We employ a uniform color map for all stars, representing the difference in time to the first observation of the corresponding night (the blue points always correspond to the start of the night and the yellow represents a different of 10.5 hours). }
		\label{Fig:asteroseismology_fwhm_corr_single_night}
		\end{figure*}

		In Section \ref{Sec:trend_introduction} we focused our analysis on a asteroseismology dataset, containing  \espresso{} data of 2 stars (\epsIndi{} and HD40307) and \HARPS{} observations of another star (HD160691). In here we present, in Figure \ref{Fig:asteroseismology_fwhm_corr_single_night}, a scatter plot between the RV residuals and the line-shape metrics.	 This comparison must, however, be interpreted carefully, as the variation of both the line-shape metrics and the RV residuals have different amplitudes in the different stars.

		Across the three stars, we find no correlations with the FWHM and bisector span. As for the contrast, we find the same lack of correlations on \epsIndi{} and \textit{HD160691} data. Strangely, that is not the case for \textit{HD40307} as we find a clear correlation between variation in contrast and RV residuals, with a correlation coefficient of 0.62. A closer look at this star reveals that the seeing gradually worsened during the night which, in turn, leads to a correlation between contrast and time. As the residuals that we measure are also correlated with time, we are left with a correlation between RV residuals and contrast.

		Lastly, it is important to note that the results of this analysis does not paint a full picture of line variability. By definition, the CCF metrics are only sensitive to variations in the lines that are present on the CCF mask. It is still possible that we are being affected by variability on the lines that are not present on the mask and, consequently, we find no clear correlations. 
	
	\section{Slopes measured in the large-sample analysis} \label{App:slope_of_large_sample}
		In Table \ref{Tab:largesample_slope_list} we present the slope of the first-degree polynomials that were adjusted to the large sample of stars that we selected, sorted by their absolute value. The slope was adjusted through a least-squares fit, using the inverse variance as weights\footnote{As implemented in \textit{numpy}: \url{https://numpy.org/doc/stable/reference/generated/numpy.polyfit.html}}. The uncertainty associated with this measurement was estimated through the diagonal of the covariance matrix, ignoring cross-correlations between parameters.
				
		\begin{table}[H]
			\centering
			\caption{\label{Tab:largesample_slope_list}Slope (S) and associated uncertainty ($\sigma_{S}$) that was adjusted to the residuals between CCF and \sbart{} RVs, when using the large sample of stars described in Section \ref{Sec:dset_overview_large_sample}}
			\begin{tabular}{c|ccc} \hline \hline
				Star     & Night      & S [\metersecondhour]  & $\sigma_{S}$ [\metersecondhour]               \\ \hline
				HD106315 & 2021-04-29 & -53.69 &    0.32    \\
				HAT-P-26 & 2021-04-09 & -27.04 &    0.55    \\
				WASP-20  & 2021-11-03 & -11.63 &    0.98    \\
				WASP-178 & 2021-07-09 & -11.33 &    1.36    \\
				WASP-54  & 2021-04-06 & -10.7  &    0.85    \\
				HD40307  & 2018-12-26 & -9.07  &    0.06    \\
				WASP-31  & 2022-01-26 & -8.19  &    0.75    \\
				Kelt-14  & 2021-01-18 & -7.83  &    0.17    \\
				HD40307  & 2018-12-27 & -7.28  &    0.06    \\
				TOI-132  & 2021-08-13 & -6.96  &    0.21    \\
				HAT-P-26 & 2021-03-23 & -6.88  &    0.15    \\
				WASP-12  & 2021-12-30 & -5.71  &    0.56    \\
				HD40307  & 2018-12-24 & -5.53  &    0.05    \\
				HD40307  & 2018-12-25 & -5.49  &    0.06    \\
				TOI-824  & 2022-06-28 & -5.44  &    0.25    \\
				WASP-156 & 2021-10-11 & -5.14  &    0.21    \\
				HD40307  & 2018-12-23 & -4.89  &    0.02    \\
				WASP-34  & 2021-02-25 & -4.32  &    0.16    \\
				WASP-126 & 2021-10-23 & -4.21  &    0.56    \\
				TOI-132  & 2022-09-18 & -3.75  &    0.16    \\
				WASP-62  & 2022-01-16 & -3.74  &    0.36    \\
				WASP-12  & 2021-11-25 & -3.72  &    0.34    \\
				Kelt-14  & 2021-02-11 & -3.59  &    0.16    \\
				WASP-62  & 2021-12-25 & -3.42  &    0.37    \\
				WASP-126 & 2019-11-19 & -3.06  &    0.2     \\
				WASP-34  & 2021-04-18 & -2.82  &    0.12    \\
				 HD3167  & 2019-10-08 & -2.19  &    0.19    \\
				WASP-54  & 2021-05-13 & -1.03  &    0.15    \\
				HD209458 & 2019-09-10 & -1.02  &    0.06    \\
				 55CncA  & 2020-02-04 & -0.75  &    0.02    \\
				HD209458 & 2019-07-19 & -0.74  &    0.05    \\
				 HD3167  & 2022-11-02 &  -0.6  &    0.08    \\
				\epsIndi & 2022-09-09 &  -0.4  &    0.01    \\
				\epsIndi & 2022-09-08 & -0.38  &    0.01    \\
				\epsIndi & 2022-09-10 & -0.36  &    0.01    \\
				\epsIndi & 2022-09-07 & -0.34  &    0.01    \\
				\epsIndi & 2022-09-06 & -0.34  &    0.01    \\
				\epsIndi & 2022-09-05 & -0.32  &    0.01    \\ \hline
			\end{tabular}
		\end{table}
		
		This analysis reveals the presence of large differences in the retrieved RV slope between  different targets and, on top of that, between different nights of the same target (.e.g., as seen in HD40307 or HAT-P-26).

		\section{Searching for correlations with weather conditions} \label{App:weather_correlation}

\begin{figure}[h!]
			\centering
			\resizebox{\hsize}{!}{\includegraphics{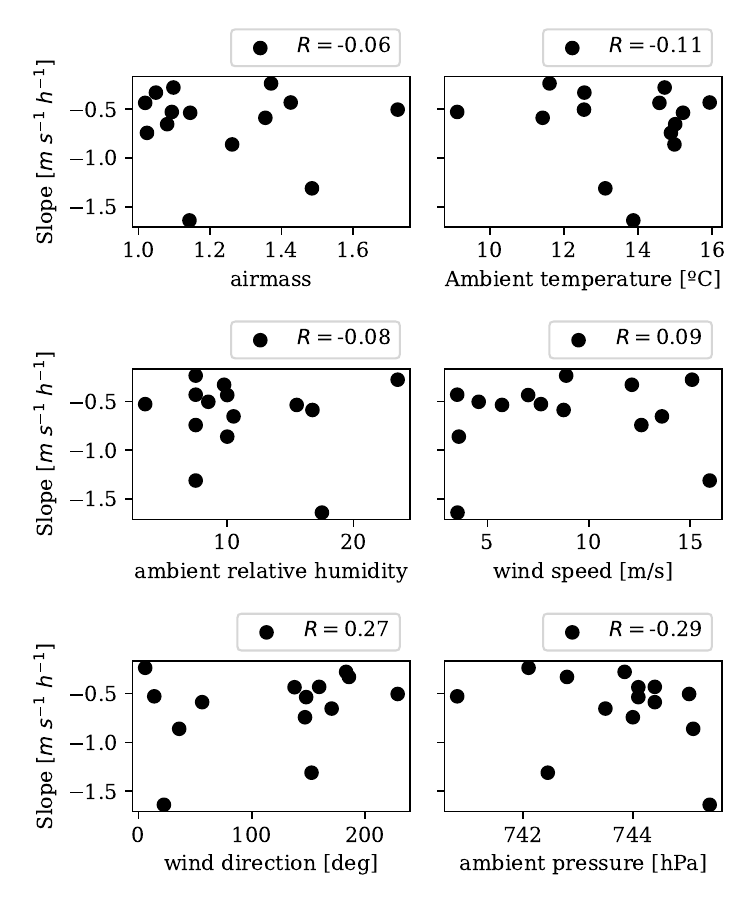}}
			\caption{Correlation between the slopes measured in Figure \ref{Fig:tauceti_single_nights} and the median value of meteorological metrics collected by ESO for the \tauCeti{} observations. In the cases where there are measurements at the start and end of the observation, we report the value at the beginning.}
			\label{Fig:tauceti_correlations}
		\end{figure}

		In this Appendix we explore a possible correlation between the meteorological conditions and the \tauCeti{} RVs extracted with the nightly templates, as presented in Section \ref{Sect:tauceti_nightly}. From the headers of the \espresso{} files we collect  the measurements of: ambient temperature, relative humidity, and pressure; airmass; wind speed and direction. For ease of analysis we plot, in Figure \ref{Fig:tauceti_correlations}, the slope and the median value of each of those metrics. 

		Unexpectedly, we fail to find a meaningful correlation with any of the aforementioned metrics. This dampens our explanation of telluric features being the root source of this contamination. If the drift was indeed caused by micro-telluric features that are bypass our mask we should expect a correlation with the metrics that influence their depth.

		\section{Impact of the BERV separation on the extraction} \label{Sec:berv_separation_impact}
		
			Lastly, we evaluate if this systematic contamination is coherent at longer time-scales or if it exists only within a small time window. In order to do so we construct stellar templates from individual nights and then use them to extract RVs. This process is applied to the two \espresso{} asteroseismology targets (\epsIndi{} and \textit{HD40307}) and for \tauCeti{} observations. The former dataset provides a dense coverage of a few consecutive nights, and the later provides a larger BERV coverage, albeit with a sparser sampling within any given night.

\begin{figure}[h!]
				\centering
				\resizebox{\hsize}{!}{\includegraphics{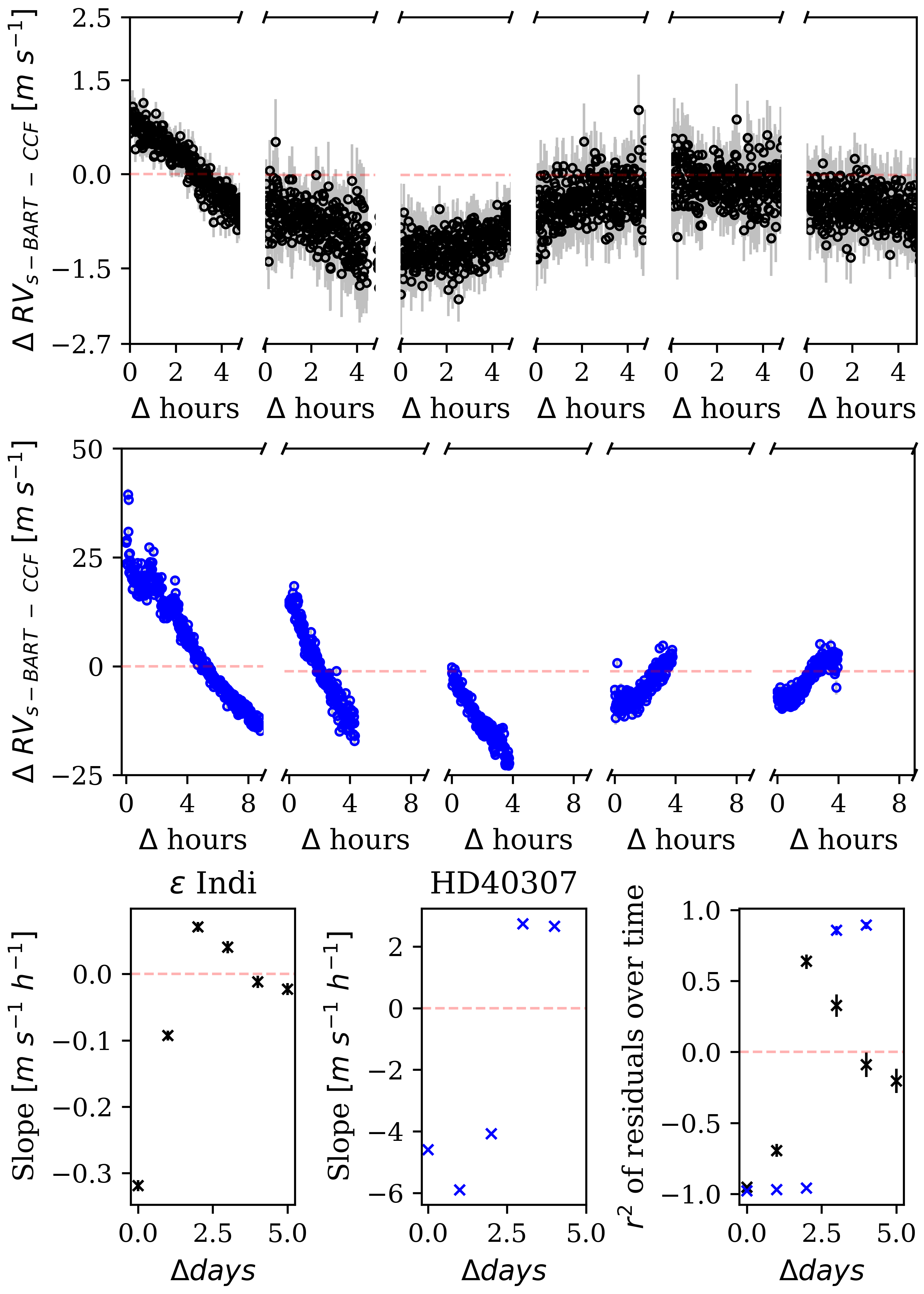}}
				\caption{RV residuals between \sbart{} and CCF RVs for the \espresso{} observations of \epsIndi{} (in black, \textbf{top row}) and \textit{HD40307} (in blue, \textbf{middle row}), divided in the different consecutive nights. The data from the first night of observations was used to construct the stellar template that was then applied to the other nights; \textbf{Bottom row:} Slope of a first degree polynomial that fitted to the RV residuals of \epsIndi{} (left)  and \textit{HD40307} (middle). The right-most column presents the \textit{Pearson} correlation coefficient of the residuals over time, with the error bars representing a 90\% confidence interval.}
				\label{fig:astero_template_grid}
			\end{figure}

			\paragraph{\textbf{Asteroseismology observations:}}
			To have a better understanding of the relation between the bias and BERV, we select the high-SNR observations of \epsIndi{} and the comparatively lower-SNR observations of \textit{HD40307}. For each star, we construct a template from the data collected on their first night and use it to extract RVs from all others, maximizing the BERV separation between the stellar template and the last observation of the dataset. Figure \ref{fig:astero_template_grid} presents the results from this comparison, clearly showing that the systematic bias disappears as the temporal distance between data and model increases. If we start by focusing on the top row (data from \epsIndi) we find that the decrease of the peak-to-peak amplitude of the residuals is accompanied by a slight increase in the dispersion. A comparison between the first two nights shows a decrease in the structure of the residuals. Strangely, the application to the third night leads to a slight positive trend and, after that point, it converges to values close to zero. The behaviour of \textit{HD40307} (middle row) differs from the one found on \epsIndi{}, as the change to a positive trend only occurs one day later (on the 4th night). Furthermore, the convergence towards "flat" residuals is slower, as the trend persists in all cases. However, the peak-to-peak impact of this systematic bias is larger on this star, in comparison with \epsIndi.

			These conclusions are summarized in the bottom panels of Figure \ref{fig:astero_template_grid}, which shows the RV slope as a function of the separation between the template and the data, alongside with \textit{Pearson}'s correlation coefficient (which is summarized in Tab. \ref{Tab:first_night_template_correlations}). One possible explanation for the difference in behavior between the two cases is that the BERV variation for \epsIndi{} is such that we there is no overlap between two consecutive nights. However, the long duration of the \textit{HD40307} nights of observations allow a significant overlap in BERV between different days.

			\begingroup 
			\renewcommand{\arraystretch}{1.5} 
			\begin{table}[h]
				\caption{\textit{Pearson}'s correlation coefficient -- R -- and 90\% confidence interval of the RV residuals over time for \epsIndi{} and \textit{HD40307} observations when applying a template constructed from the first night available to us.}
				\label{Tab:first_night_template_correlations}
				\centering
				\begin{tabular}{ccc}\hline \hline
					$\Delta$ days  & $R$ of \epsIndi{} & $R$ of \textit{HD40307} \\ \hline
					0      & $-0.95^{0.01}_{-0.01}$  & $-0.98^{0.00}_{-0.00}$  \\
					1      & $-0.70^{0.05}_{-0.04}$  & $-0.97^{0.01}_{-0.01}$  \\
					2      & $ 0.64^{0.05}_{-0.05}$  & $-0.96^{0.01}_{-0.01}$  \\
					3      & $ 0.33^{0.08}_{-0.08}$  & $ 0.86^{0.03}_{-0.04}$  \\
					4      & $-0.09^{0.09}_{-0.09}$  & $ 0.89^{0.02}_{-0.03}$  \\
					5      & $-0.20^{0.09}_{-0.08}$  &            --             \\ \hline
				\end{tabular}
				\tablefoot{ The correlation coefficient is shown as a function of the number of days between the night that was used to construct the stellar template and the night from which we extract the RVs (refer to Appendix \ref{Sec:berv_separation_impact} for details).}
			\end{table}
			\endgroup

			\paragraph{\textbf{\tauCeti{} observations:}} 

			The analysis of \tauCeti{} data allows us to expand on this analysis over a longer timescale. We select the night of 4th of December, 2022 as a baseline. Then, we select all nights that were observed in a window of 50 days before and after it. In total, we collect 17 nights (including the baseline one) that are then used to construct single-night stellar templates. We use each of those templates to compute \sbart{} RVs for the baseline night.
			
\begin{figure}[h]
				\centering
				\resizebox{\hsize}{!}{\includegraphics{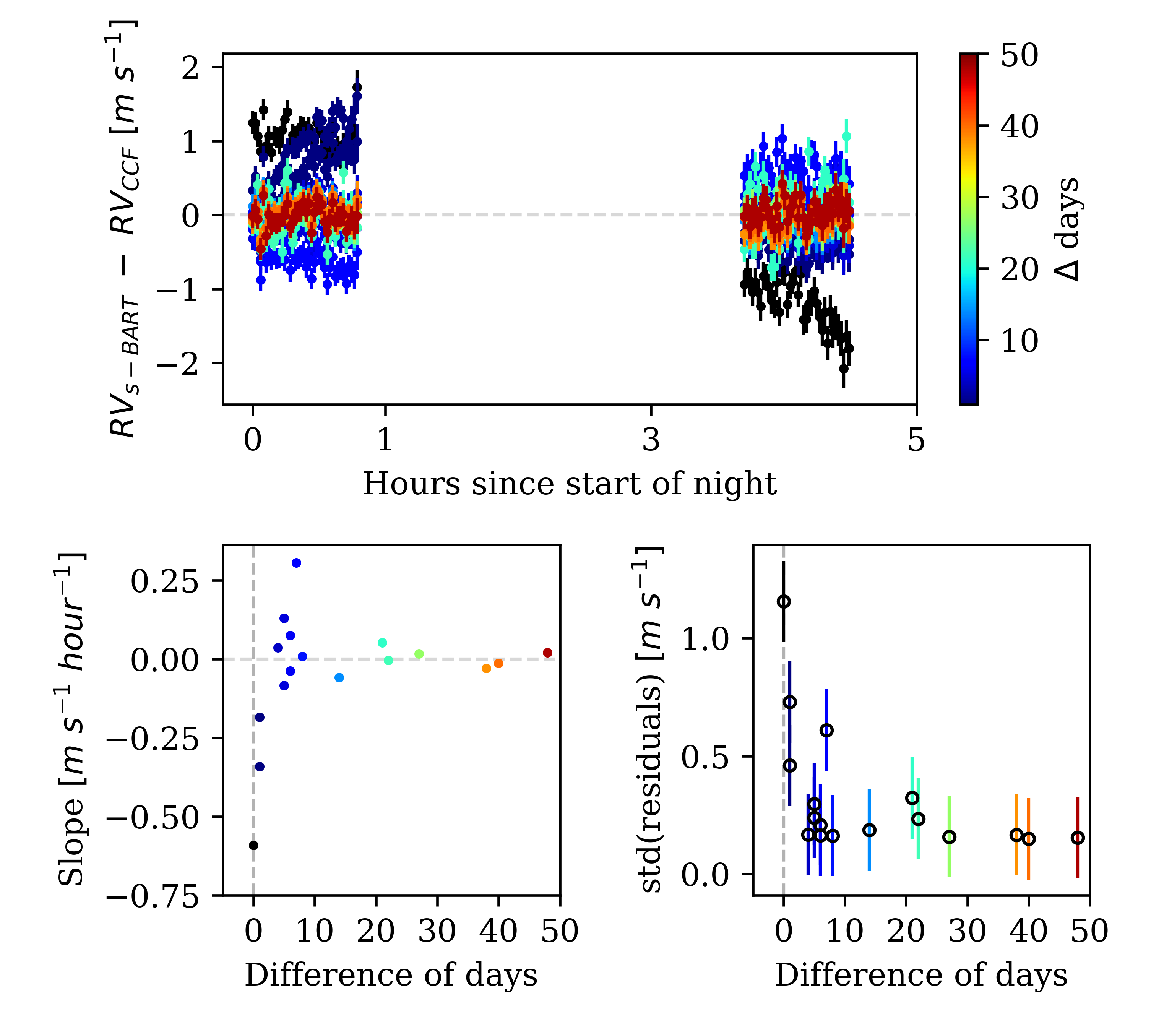}}
				\caption{\textbf{Top panel}: Residuals between the \sbart{} and CCF RVs of \tauCeti, color-coded by the difference in days between the baseline night and the night from which the template was constructed. \textbf{Bottom panel:} Evolution of two metrics with the the distance (in days) between template and data: slope of a first degree polynomial (left) and the standard deviation of the RV residuals (right).}
				\label{fig:TauCeti_year_of_data}
			\end{figure}

			Figure \ref{fig:TauCeti_year_of_data} presents the evolution of the residual slope as a function of temporal separation between observation and template. The slope converges to zero after 20 days of difference between data and template and takes a positive value in some cases. The scatter in the residuals also converges to zero after about 10 days of separation.
			
			Finally, it is important to note that this analysis of the three stars is not enough for us to be fully certain that a given temporal separation is enough to mitigate most of the effect of this RV contamination. We have previously seen that the amplitude of this parasitic effect has a large variation across different stars (with a more pronounced effect in lower-SNR regimes). In the analysis of these 3 stars we have also seen that the decrease of the RV slope occurs at different rates for different stars.

	\section{The effect of telluric correction} \label{App:TCtellcorr}
			
		The latest release of \espresso{} pipeline has introduced telluric correction for water features, following the methodology of \citet{allartAutomaticModelbasedTelluric2022}. As we discussed throughout the paper, the hypothesis of telluric contamination would imply that we have micro-telluric features that are not accounted for in our synthetic transmittance profile. One avenue to test this is to apply the same \sbart{} recipe -- that we used in this work -- to the telluric-corrected spectra. It is important to note that this new RV extraction will apply the telluric-masking on the telluric-corrected spectra. This means that the only difference at the level of the stellar spectra is in the micro-telluric features that bypass our binary mask.

\begin{figure}[h]
			\centering
			\resizebox{\hsize}{!}{\includegraphics{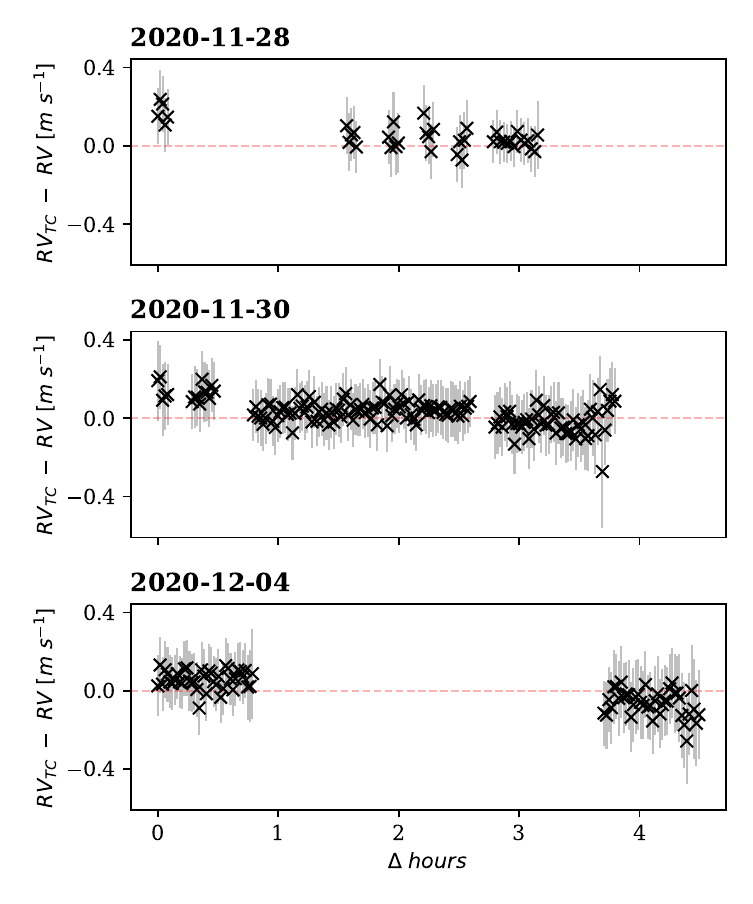}}
			\caption{Comparison of the \tauCeti{} RV time-series extracted from telluric corrected spectra for three nights explored in this paper.}
			\label{fig:TauCetiTC}
		\end{figure}

		In Figure \ref{fig:TauCetiTC} we present the residuals between the two \sbart{} RVs, showing an agreement at the 1-$\sigma$ level. There is a slight difference between the two RV time-series, but it is not significant. A follow-up analysis with telluric-correction on other molecules will be done after it is implemented in \espresso's DRS.

		Since we do not find significant differences in the telluric-corrected spectra, it is even more unlikely that we are indeed missing telluric lines from our model.

	\section{Other possible sources of RV biases} \label{App:othercauses}
  
		In this Appendix we postulate on other effects that could introduce trends within datasets that are collected within a single night:

		\begin{enumerate}
			\item A short-scale variability from the spectral continuum, that is biasing the stellar RVs. One of the main differences between CCF and TM algorithms is the regions of the stellar spectra that they use. In template-matching algorithms the stellar model will contain both spectral lines and the continuum, whilst on the CCF the model only contains the stellar lines (usually weighted by their RV information). Within \sbart{} we assume that the continuum of a given observation can have a linear difference to the stellar model. In other words, we marginalize over the hyper-parameters of a first-degree polynomial that describes the flux difference. If we have an higher-order variation on the flux levels over a night, they would not be accounted for. However, it is not clear to us why a change in the continuum level would translate into a multi-meter per second bias on RV time-series.
			\item A contamination between fiber A and fiber B of our instrument. In nominal operations\footnote{For \espresso{} and \HARPS{}, but the discussion is also applicable to other spectrographs.}, fiber A collects light from the science target, whilst fiber B will either be placed on the sky (to allow for the subtraction of its effect) or on a calibration light-source\footnote{We refer the reader to the DRS manual \citep{eso_espresso_2023} for further information}. It is not clear why such an effect would not appear on CCF RV and, on top of that, we find this bias on data collected with the two configurations of fiber B, i.e., both for simultaneous calibration or sky. This rules out the possible, but unlikely, contaminations between the two fibers.
			\item The last effect that we can think of is a possible injection of a systematic effect of the stellar spectra from the calibration frames that are used by the instrument's official pipeline. When the DRS is applied to the raw images, it selects the calibration frames closest to each image. This means that the DRS will reduce every observation (of the same night) using the same set of calibration frames, which is not the case in observations taken in different nights. A possible deviation of the instrumental properties, in comparison with the calibration frame, could be leading to an intra-night evolution of the stellar spectra. A preliminary analysis of some of the calibration frames (flat-fields and darks) revealed no obvious artifacts, but we leave a full analysis of those data products for future work. Furthermore, we find this effect not only on \espresso{} and \HARPS{}, but we are also aware of the same systematic bias on data from other high-resolution spectrographs\footnote{Through personal communication.}. 
		\end{enumerate}

\end{appendix}

\end{document}